\documentclass[twocolumn]{aastex63}

\pdfoutput=1
\NewPageAfterKeywords

\usepackage{enumitem}
\usepackage{physics}
\usepackage{multirow}

\hypersetup{pdfnewwindow=true,
     colorlinks=true,
     linkcolor=blue,
     citecolor=blue,
	 final=true}
	 
\usepackage{etoolbox}
\makeatletter
\patchcmd{\NAT@citex}
  {\@citea\NAT@hyper@{%
     \NAT@nmfmt{\NAT@nm}%
     \hyper@natlinkbreak{\NAT@aysep\NAT@spacechar}{\@citeb\@extra@b@citeb}%
     \NAT@date}}
  {\@citea\NAT@nmfmt{\NAT@nm}%
   \NAT@aysep\NAT@spacechar\NAT@hyper@{\NAT@date}}{}{}
\patchcmd{\NAT@citex}
  {\@citea\NAT@hyper@{%
     \NAT@nmfmt{\NAT@nm}%
     \hyper@natlinkbreak{\NAT@spacechar\NAT@@open\if*#1*\else#1\NAT@spacechar\fi}%
       {\@citeb\@extra@b@citeb}%
     \NAT@date}}
  {\@citea\NAT@nmfmt{\NAT@nm}%
   \NAT@spacechar\NAT@@open\if*#1*\else#1\NAT@spacechar\fi\NAT@hyper@{\NAT@date}}
  {}{}
\makeatother

\newcommand{\hii}{${\rm H}\,\textsc{ii}$}

\newcommand{\cii}{$[{\rm C}\,\textsc{ii}]$}
\newcommand{\oiii}{$[{\rm O}\,\textsc{iii}]$}
\newcommand{\Lcii}{$L_{[{\rm C}\,\textsc{ii}]}$}
\newcommand{\Loiii}{$L_{[{\rm O}\,\textsc{iii}]}$}
\newcommand{\Sigcii}{$\Sigma_{[{\rm C}\,\textsc{ii}]}$}
\newcommand{\Sigoiii}{$\Sigma_{[{\rm O}\,\textsc{iii}]}$}
\newcommand{\Sigsfr}{$\Sigma_{\rm SFR}$}
\newcommand{\zD}{A1689-zD1}

\begin{document}

\title{\large ALMA reveals extended cool gas and hot ionized outflows \\ in a typical star-forming galaxy at $z=7.13$}

\shortauthors{Akins et al.}
\shorttitle{Extended \cii\ in \zD}

\correspondingauthor{Hollis B. Akins} 
\email{hollis.akins@gmail.com}

\author[0000-0003-3596-8794]{Hollis B. Akins}
\affiliation{Department of Physics, Grinnell College, 1116 Eighth Ave., Grinnell, IA 50112, USA}

\author[0000-0001-7201-5066]{Seiji Fujimoto}
\affiliation{Cosmic Dawn Center (DAWN), Jagtvej 128, DK2200 Copenhagen N, Denmark}
\affiliation{Niels Bohr Institute, University of Copenhagen, Lyngbyvej 2, DK2100 Copenhagen \O, Denmark}

\author[0000-0002-0496-1656]{Kristian Finlator}
\affiliation{Department of Astronomy, New Mexico State University, Las Cruces, New Mexico, USA}
\affiliation{Cosmic Dawn Center (DAWN), Jagtvej 128, DK2200 Copenhagen N, Denmark}

\author[0000-0002-4465-8264]{Darach Watson}
\affiliation{Cosmic Dawn Center (DAWN), Jagtvej 128, DK2200 Copenhagen N, Denmark}
\affiliation{Niels Bohr Institute, University of Copenhagen, Lyngbyvej 2, DK2100 Copenhagen \O, Denmark}

\author[0000-0002-7821-8873]{Kirsten K. Knudsen}
\affiliation{Department of Space, Earth and Environment, Chalmers University of Technology, Onsala Space Observatory, SE-43992 Onsala, Sweden}

\author[0000-0001-5492-1049]{Johan Richard}
\affiliation{Univ Lyon, Univ Lyon1, ENS de Lyon, CNRS, Centre de Recherche Astrophysique de Lyon UMR5574, 69230 Saint-Genis-Laval, France}

\author[0000-0002-5268-2221]{Tom J. L. C. Bakx}
\affiliation{Division of Particle and Astrophysical Science, Graduate School of Science, Nagoya University, Aichi 464-8602, Japan}
\affiliation{National Astronomical Observatory of Japan, 2-21-1 Osawa, Mitaka, Tokyo 181-8588, Japan}

\author[0000-0002-0898-4038]{Takuya Hashimoto} 
\affiliation{Tomonaga Center for the History of the Universe (TCHoU), Faculty of Pure and Applied Science, University of Tsukuba, Ibaraki, 305-8571, Japan}

\author[0000-0002-7779-8677]{Akio K.~Inoue}
\affiliation{Department of Physics, School of Advanced Science and Engineering, Faculty of Science and Engineering, Waseda University, 3-4-1 Okubo, Shinjuku, Tokyo 169-8555, Japan}
\affiliation{Waseda Research Institute for Science and Engineering, Faculty of Science and Engineering, Waseda University, 3-4-1 Okubo, Shinjuku, Tokyo 169-8555, Japan}

\author[0000-0003-3278-2484]{Hiroshi Matsuo}
\affiliation{National Astronomical Observatory of Japan, 2-21-1 Osawa, Mitaka, Tokyo 181-8588, Japan}
\affiliation{Graduate University for Advanced Studies (SOKENDAI), 2-21-1 Osawa, Mitaka, Tokyo 181-8588, Japan}

\author[0000-0001-9033-4140]{Micha\l{} J. Micha\l{}owski}
\affiliation{Astronomical Observatory Institute, Faculty of Physics, Adam Mickiewicz University, ul. S\l{}oneczna 36, 60-286 Pozna\'n, Poland}

\author[0000-0003-4807-8117]{Yoichi Tamura}
\affiliation{Division of Particle and Astrophysical Science, Graduate School of Science, Nagoya University, Aichi 464-8602, Japan}

\received{November 22, 2021}
\revised{June 9, 2022}
\accepted{June 14, 2022}
\submitjournal{ApJ}

\begin{abstract} 
	We present spatially-resolved morphological properties of \cii~$158~\mu$m, \oiii~$88~\mu$m, dust, and rest-frame ultraviolet (UV) continuum emission for \zD, a strongly lensed, sub-L* galaxy at $z=7.13$, by utilizing deep Atacama Large Millimeter/submillimeter Array (ALMA) and {\it Hubble Space Telescope} ({\it HST}) observations.
	While the \oiii\ line and UV continuum are compact, the \cii\ line is extended up to a radius of $r \sim 12$ kpc.  
	Using multi-band rest-frame far-infrared (FIR) continuum data ranging from 52 -- 400~$\mu$m, we find an average dust temperature and emissivity index of $T_{\rm dust} = 41^{+17}_{-14}$ K and $\beta = 1.7^{+1.1}_{-0.7}$, respectively, across the galaxy. 
	We find slight differences in the dust continuum profiles at different wavelengths, which may indicate that the dust temperature decreases with distance. 
	We map the star-formation rate (SFR) via IR and UV luminosities and determine a total SFR of $37\pm 1~M_\odot~{\rm yr}^{-1}$ with an obscured fraction of $87\%$.
	While the \oiii\ line is a good tracer of the SFR, the \cii\ line shows deviation from the local \Lcii-SFR relations in the outskirts of the galaxy. 
	Finally, we observe a clear difference in the line profile between \cii\ and \oiii, with significant residuals ($\sim 5\sigma$) in the \oiii\ line spectrum after subtracting a single Gaussian model. 
	This suggests a possible origin of the extended \cii\ structure from the cooling of hot ionized outflows. 
	The extended  \cii\ and high-velocity \oiii\ emission may both contribute in part to the high \Loiii/\Lcii\ ratios recently reported in $z>6$ galaxies.
	\end{abstract}

\keywords{\small {\it Unified Astronomy Thesaurus concepts:} \href{http://astrothesaurus.org/uat/594}{Galaxy evolution (594)}; \href{http://astrothesaurus.org/uat/595}{Galaxy formation (595)}; \href{http://astrothesaurus.org/uat/734}{High-redshift galaxies (734)}; \href{http://astrothesaurus.org/uat/847}{Interstellar medium (847)}; \href{http://astrothesaurus.org/uat/1879}{Circumgalactic medium (1879)}}

\section{Introduction}\label{sec:intro}

The formation and evolution of galaxies across cosmic time is shaped by the ``baryon cycle,'' namely the complex interplay between gas accretion, star formation, stellar and active galactic nucleus (AGN) feedback, and gas outflows \citep[e.g.][]{dave_analytic_2012, angles-alcazar_cosmic_2017c, tumlinson_circumgalactic_2017}.
These various processes serve to transfer gas between the interstellar medium (ISM) and the surrounding circumgalactic medium (CGM).
Key to understanding the role of the baryon cycle in the formation of the first galaxies are studies investigating the morphology and structure of the ISM and CGM at high redshift, where robust observational constrains are rare.

At high redshift, far-infrared (FIR) fine-structure lines are typically used to trace the ISM and the galaxies' star-formation activity.
In particular, the \cii\ $^2P_{3/2} \to {^2P_{1/2}}$ transition at $157.74~\mu$m (hereafter \cii), with its low ionization potential of 11.2 eV (compared to 13.6 eV of Hydrogen) is a dominant coolant of the neutral ISM \citep{stacey_cii_1991,wolfire_neutral_2003}, and dense photodissociation regions \citep[PDRs;][]{hollenbach_ism_1999} associated with giant molecular clouds.
The \cii\ line, arising from singly ionized carbon, has been used to probe ISM properties \citep[e.g.][]{capak_galaxies_2015, pentericci_tracing_2016, knudsen_cii_2016,carniani_extended_2017,bakx_alma_2020,matthee_resolved_2020} 
and kinematics \citep[e.g.][]{jones_dynamical_2017, smit_rotation_2018, rizzo_dynamical_2020, rizzo_dynamical_2021} in dozens of high redshift galaxies. 
While the \cii\ line traces multiple phases of the ISM, it has been shown to primarily trace cool, neutral gas through PDRs \citep{cormier_herschel_2019}. 
By contrast, the \oiii\ $^3P_1 \to ^3P_0$ transition at $88.4~\mu$m (hereafter \oiii) primarily traces hot, diffuse, ionized gas, as the ionization potential of ${\rm O}^+$ ($35.1$ eV) is significantly higher than that of Hydrogen. 
Moreover, both the \cii\ and \oiii\ lines are known to trace the star-formation rate \citep[SFR;][]{kapala_survey_2014, delooze_applicability_2014, herrera-camus_ii_2015}.

The advent of Atacama Large Millimeter/submillimeter Array (ALMA) opens our observational window to detect both the \cii\ and \oiii\ lines at $z > 4$, and the number of detections risen rapidly in recent years \citep[see][]{hodge_highredshift_2020}. 
Of particular relevance, recent ALMA stacking studies have discovered extended \cii\ structures up to $r\sim 10$--$15$ kpc around high redshift galaxies \citep{fujimoto_first_2019, ginolfi_alpinealma_2020}.  
The extended structure, called the ``\cii\ halo'' \citep{fujimoto_first_2019}, shows a $\sim 5$ times larger effective radius than the central galaxy as seen in the UV and a potential association with the Ly$\alpha$ halo \citep[e.g.][]{leclercq_muse_2017}. 
Similar extended \cii\ structures have also been identified in observations of individual star-forming galaxies at $z > 4$ \citep{fujimoto_alpinealma_2020, herrera-camus_kiloparsec_2021a} and massive quasar host galaxies at $z > 6$ \citetext{\citealp{maiolino_evidence_2012,cicone_very_2015}; \citealp[cf.][]{meyer_physical_2022}}.

In several cases, observations of the extended \cii\ structure have additionally detected the ``broad-wing'' feature in the \cii\ spectrum \citep{gallerani_alma_2018, ginolfi_alpinealma_2020, herrera-camus_kiloparsec_2021a}, characteristic of ongoing high-velocity outflows, which have been hypothesized to drive carbon-enriched gas out into the CGM.
The existence of extended \cii\ emission is not fully reproduced by the current galaxy formation models \citep{fujimoto_first_2019, arata_galaxy_2020,pallottini_deep_2019,katz_probing_2019a,katz_probing_2019b,katz_nature_2022} suggesting that there may be missing physics in simulations at the Epoch of Reionization \citep{pizzati_outflows_2020}.

\citet{fujimoto_first_2019} discuss several possible scenarios for what physical processes may power the extended \cii\ emission, including outflows, satellite galaxies (in which \cii\ emission traces their SFR), and photoionization via radiation from the central galaxy. 
\citet{ginolfi_alpinealma_2020} and \citet{fujimoto_alpinealma_2020}  find correlations between galaxy SFR and the detection of the broad-wing feature and \cii\ extension, suggestive of the SF-driven outflow scenario. 
However, this result may be biased by the signal to noise, as objects with higher SFR are likely brighter, making faint components in the spectra easier to detect \citep{ginolfi_alpinealma_2020,fujimoto_alpinealma_2020}.
\citet{ginolfi_alpinealma_2020a} introduce an additional scenario, tidal disruption, which can produce \cii\ emission on CGM scales through turbulence and shocks through gravitational interactions.
Moreover, while the existence of the extended \cii\ structure implies the presence of dust in the CGM, high redshift ALMA observations have generally found a compact dust continuum \citep{fujimoto_first_2019, ginolfi_alpinealma_2020}.
In contrast, observations of local galaxies suggests that dust entrained by galactic winds can reach $\sim 10$ kpc in scale \citep{kaneda_akari_2009, melendez_exploring_2015,mccormick_exploring_2018, yoon_exploring_2021} and theory suggests that stellar feedback is efficient at driving dust-enriched, multiphase winds \citep{kannan_dust_2021}.
ALMA observations of quasar host galaxies at $z \gtrsim 6$ have found \cii\ and dust emission extended out to $r\sim 10$ kpc, and interpret both as tracing the star-forming ISM of these massive galaxies \citep{novak_no_2020}.
A proper determination of the physical processes producing the extended \cii\ structure in line with the outflow thus depends on our ability to measure, with sufficient depth, the distribution of neutral/ionized gas and dust continuum emission. 
Simultaneous ALMA observations of the \cii\ and \oiii\ fine-structure lines, and their underlying continua, are well-suited to accomplish this task.

In this work, we utilize deep ALMA observations of the galaxy \zD, a strongly lensed, sub-L* galaxy at $z=7.13$ \citep{bradley_discovery_2008, watson_dusty_2015, knudsen_merger_2017}.
The high magnification ($\mu \sim 9.3$) of this galaxy makes it an ideal target for a spatially-resolved investigation of the ISM in the abundant population of sub-L* galaxies in the Epoch of Reionization, and indeed, recent high-resolution follow-up observations have successfully detected bright \cii~$158~\mu$m and \oiii~$88~\mu$m emission lines (Knudsen et al. in prep). 
Moreover, rich FIR continuum data has also been taken in bands 3, 6, 7, 8, and 9, providing valuable constraints on the FIR luminosity and dust temperature \citep{bakx_accurate_2021}.
The availability of deep, high resolution, multiwavelength ALMA data make A1689-zD1 the ideal target to address the aforementioned issues and examine the extended \cii\ emission around high redshift galaxies.

The paper is structured as follows: 
In Section~\ref{sec:observations} we describe the data reduction steps for both ALMA and {\it HST} data. 
We report the results for the spatial extent of the \cii, \oiii, UV, and FIR emission in Sections~\ref{sec:maps} and \ref{sec:morphology}. 
In Section~\ref{sec:Tdust} we spatially-resolve the dust temperature, which we use in Section~\ref{sec:SFR} to derive the spatially-resolved \Lcii-SFR and \Loiii-SFR relations for \zD. 
We present the line spectra of \cii\ and \oiii\ in Section~\ref{sec:line_profiles}. 
Finally, in Section~\ref{sec:discussion} we discuss our results, including the interpretation of the extended emission as a ``\cii\ halo'' vs.~an extended disk and the physical origin of the emission.
Throughout this work, we assume a $\Lambda$CDM cosmology with $H_0 = 67.4$ km s$^{-1}$ Mpc$^{-1}$, $\Omega_M = 0.315$, $\Omega_{\Lambda} = 0.685$ \citep{planckCollaboration_2020} and a \citet{kroupa_variation_2001} initial mass function (IMF).

\section{Data Processing}\label{sec:observations}

\begin{deluxetable*}{LCCCCCCc}
\tablecaption{Summary of ALMA continuum observations for \zD.\label{tab:data}}
\colnumbers
\tablehead{
	\colhead{\multirow{2}{*}{Band}} & \colhead{$\lambda_{\rm rest}$} & \colhead{$\lambda_{\rm min}$,~$\lambda_{\rm max}$} & \colhead{$\theta_{\rm fid}$} & \colhead{$\theta_{\rm tap}$} & \colhead{$\sigma_{\rm fid}$} & \colhead{$\sigma_{\rm tap}$} & \colhead{\multirow{2}{*}{Reference}} \\[-0.7em]
	\colhead{~} & \colhead{\footnotesize [$\mu$m]} & \colhead{\footnotesize [$\mu$m]} & \colhead{\footnotesize[arcsec]} & \colhead{\footnotesize[arcsec]} & \colhead{\footnotesize[$\mu$Jy beam$^{-1}$]} & \colhead{\footnotesize[$\mu$Jy beam$^{-1}$]} & \colhead{~}\\[-2.5em]}
\startdata
3 & 402.5 & 375.8,~429.2 & 0.55 \times 0.46 &    -    & 10.4 &    -    & Knudsen et al.~(in prep) \\
6^\dagger & 163.0 & 156.2,~169.9 & 0.44 \times 0.40 & 0.82 \times 0.73 & 5.3 & 8.3 & Knudsen et al.~(in prep)$^{\ddagger}$ \\
7 & 107.2 & 104.7,~109.7 & 0.62 \times 0.58 & 0.78 \times 0.74 & 44.9 & 47.1 & \citet{knudsen_merger_2017} \\
8 & 89.5 & 87.8,~91.1 & 0.46 \times 0.42 & 0.72 \times 0.69 & 24.5 & 28.7 & Knudsen et al.~(in prep)$^{\ddagger}$ \\
9 & 53.0 & 52.6,~53.4 & 0.52 \times 0.43 & 0.67 \times 0.61 & 181.2 & 205.0 & \citet{bakx_accurate_2021}
\enddata
\tablecomments{Columns: (1) ALMA band, (2-3) Central wavelength and range, (4-5) Beam size, for ``fiducial'' maps (with no {\it uv}-taper) and ``tapered'' maps (with a {\it uv}-taper of $0.5''$), (6-7) RMS noise, (8) Reference for ALMA data.}	
\tablenotetext{\dagger}{For band 6, the ``fiducial'' and ``tapered'' maps have {\it uv}-tapers of $0.3''$ and $0.7''$, respectively, to more closely match the other bands in resolution.}
\tablenotetext{\ddagger}{Band 6 and band 8 data are also presented in \citet{wong_alma_2022}.}
\end{deluxetable*}

\subsection{ALMA Data}

We utilize ALMA observations of A1689-zD1 (R.A.~$13^{\rm h}\,11^{\rm m}\,29.9^{\rm s}$, Dec.~$-1^{\circ}\,19'\,18.7''$) in bands 3, 6, 7, 8, and 9. 
The observations in bands 6 and 8 (ALMA program IDs 2015.1.01406.S and 2017.1.00775.S, respectively) will be presented in Knudsen et al.~(in prep). 
We use archival observations for bands 7 \citep{knudsen_merger_2017}, and 9 \citep{bakx_accurate_2021}, and refer the reader to these papers for more details. 
While previous studies of \zD\ have adopted the optical redshift of $z=7.5$ determined from an apparent continuum break \citep[interpreted as the Lyman-$\alpha$;][]{watson_dusty_2015}, new observations of the \cii\ and \oiii\ lines at high S/N place the spectroscopic redshift at $z = 7.132$. 
As such, the band 6 observations were tuned to the \cii\ $158~\mu$m line at $\nu_{\rm obs} = 233.71$ GHz. 
Similarly, bands 8 and 9 were tuned to the \oiii\ $88~\mu$m, and \oiii\ $52~\mu$m lines, respectively. 
Analysis of the \oiii~$52~\mu$m line will be presented in a separate, upcoming paper. 
Band 3 observations were tuned to cover the CO(7-6) and CO(6-5) lines.

The Common Astronomy Software Applications package \citep[CASA;][]{mcmullin_CASA_2007} was used for reduction, calibration, and imaging. 
The data were first processed via the standard ALMA pipeline reduction, which was sufficient for these observations. 
All images were produced with the {\tt tclean} task in CASA using natural weighting to maximize sensitivity. 
Cleaning was done using the ``auto-multithresh" automasking algorithm \citep{kepley_automultithresh_2020} initially down to the $3\sigma$ level but expanded to a {\tt lownoisethreshold} of $1\sigma$.
In order to image both compact and extended emission, we apply the {\tt multiscale} deconvolver \citep{cornwell_multiscale_2008} with scales of 0 (delta-function) 1, and 3 times the beam size (i.e.,~the psf). 
Continuum maps were produced using the multi-frequency synthesis \citep[{\tt mfs};][]{conway_mfs_1990} mode in CASA over a frequency range identified as lacking line emission. 
To produce moment 0 maps for the \cii\ and \oiii\ lines, we first subtract the continuum using the CASA task {\tt uvcontsub}, with order 0, and produce a map using the {\tt mfs} mode.
While this mode is typically used for continuum imaging, we employ it to produce an average line intensity map, as a pseudo-moment 0 map, in order to maximize the signal-to-noise of the map prior to cleaning and ensure that we detect and deconvolve any faint and/or diffuse signals in the {\tt tclean} process.
We integrate channels between $233.4$ and $234.0$ GHz for the \cii\ map and between $416.6$ and $417.8$ GHz for the \oiii\ map. Both ranges correspond to roughly $\pm 400$ km/s from the line center.

For \cii\ and \oiii\ line intensity maps we obtain an rms of $18.8~\mu{\rm Jy}~{\rm beam}^{-1}$ and $63.0~\mu{\rm Jy}~{\rm beam}^{-1}$. 
For bands 3, 6, 7, 8, and 9 continuum maps we obtain an rms of $10.4$, $5.3$, $44.9$, $24.5$, and $181.2~\mu{\rm Jy}~{\rm beam}^{-1}$.
All maps are primary-beam corrected, though the reported rms is derived prior to primary-beam correction. 
All ALMA and {\it HST} images are aligned astrometrically using the {\tt reproject} package in Python.

With natural weighting, the synthesized beam sizes for band 6 and 8 observations are $0.30'' \times 0.28''$ and $0.46'' \times 0.41''$, respectively. 
To perform a fair comparison based on the same spatial resolution, we apply a {\it uv}-taper of $0.3''$ for the band 6 data, which increases the beam size to $0.44'' \times 0.40$.
For the analysis of extended emission (Section~\ref{sec:morphology}) we apply a {\it uv}-taper of $0.5''$ ($0.7''$) to observations in band 8 (band 6).
Table~\ref{tab:data} summarizes the beam size and depth of both the fiducial ({\it uv}-taper of $0.3''$ for band 6, untapered for other bands) and tapered ($0.7''$ for band 6, $0.5''$ for other bands) continuum data in each band.

\subsection{HST Data}

In addition to deep ALMA observations of FIR line and continuum emission, we use archival Hubble Space Telescope ({\it HST}) data to probe the rest-frame UV continuum of A1689-zD1 \citep{watson_dusty_2015}. 
We obtain images taken with Hubble WFC3/IR in the $J$ (F125W) and $H$ (F160W) bands. 
At $z = 7.13$, these bands trace the UV continuum at $1350$ \AA $< \lambda < 2050$~\AA.
These wavelengths are redder than the Lyman limit and unaffected by the Ly$\alpha$ forest, making this a good probe of the rest-frame UV continuum and the unobscured SFR.

Recent works \citep[e.g.][]{rujopakarn_vla_2016,dunlop_deep_2017,fujimoto_first_2019,tamura_detection_2019, herrera-camus_kiloparsec_2021a} have shown that there sometimes exist offsets of $0.1$-$0.3$ arcseconds between {\it HST} and ALMA astrometry, which can significantly impact results of comparisons between high-resolution observations.
The {\it HST} data we use in this work has been corrected for this astrometric offset by comparison to astrometry from the {\it Gaia} survey (Gaia Collaboration~\citeyear{gaiacollaboration_gaia_2018}), which is accurate to ALMA astrometry to the milliarcsecond scale. 
The correction shifts the {\it HST} astrometry $-0.11''$ in right ascension and $+0.06''$ in declination.

As noted in previous analysis of A1689-zD1 \citep{watson_dusty_2015, knudsen_merger_2017}, there is an additional source $\sim 1.5''$ westwards of the target that has been identified as a low-redshift ($z\sim 2$) galaxy. 
Other sources nearby \zD\ can be identified as low-$z$ interlopers galaxies by their detection in shorter-wavelength ACS/F814W observations, where \zD\ disappears due to the Lyman-break nature of the galaxy. 
We regard pixels that have S/N $>3$ in F125W+F160W and are more than $1.1''$ from \zD, that are also detected in ACS/F814W, as interlopers and replace them with randomly drawn values from the image rms noise.  
Finally, in order to present a fair comparison with the same spatial resolution between {\it HST} and ALMA images of A1689-zD1, we convolve the {\it HST} map with a Gaussian kernel constructed to match the idealized (i.e.,~Gaussian model) ALMA beam for the band 8 data. 
For the fiducial (tapered) maps, this kernel has a size of $0.44'' \times 0.39''$ ($0.70'' \times 0.67''$).

\subsection{Source Plane Reconstruction}\label{sec:spr}

To analyze the intrinsic ISM morphology of \zD, we reconstruct the ALMA and \textit{HST} images into the source plane, correcting for the lens magnification and deflection. 

The magnification of \zD\ is derived to be a factor of $\mu = 9.3\pm 0.5$ based on the mass model originally published in \citet{limousin_2007}, and refined to include spectroscopic information of multiple systems as described in \citet{bina_2016}. 
This parametric strong lensing model is constructed using the \textsc{lenstool} \citep{jullo_2007} software which provides a set of models sampling the posterior distribution of each parameter of the mass distribution, allowing us to derive the statistical error on the magnification. 
The uncertainty of $0.025$ dex of the magnification at the location of the source was also confirmed independently as described in \citet{watson_dusty_2015}, and the magnification is quite uniform across the source, with variations around 5\% (see Figure~\ref{fig:mag_map} in Appendix~\ref{appendix:A}).
We show source plane reconstructed maps for each emission in Figure~\ref{fig:maps_uv}; we employ these source plane maps when measuring distance scales (i.e.~Sections~\ref{sec:radial} and \ref{sec:gini}).
This is because the asymmetrical deflection of the source confuses a simple relation between angular and physical scale in the image plane---the full reconstruction is necessary to assess the size/shape of the source. 

When quoting the luminosity or flux from the source, we employ the notation that, for example, $L_{\rm IR}$ represents the observed IR luminosity and $L_{\rm IR}\,\mu^{-1}$ represents the intrinsic IR luminosity, corrected for magnification.

\begin{figure*}
	\centering
	\includegraphics[width=\linewidth]{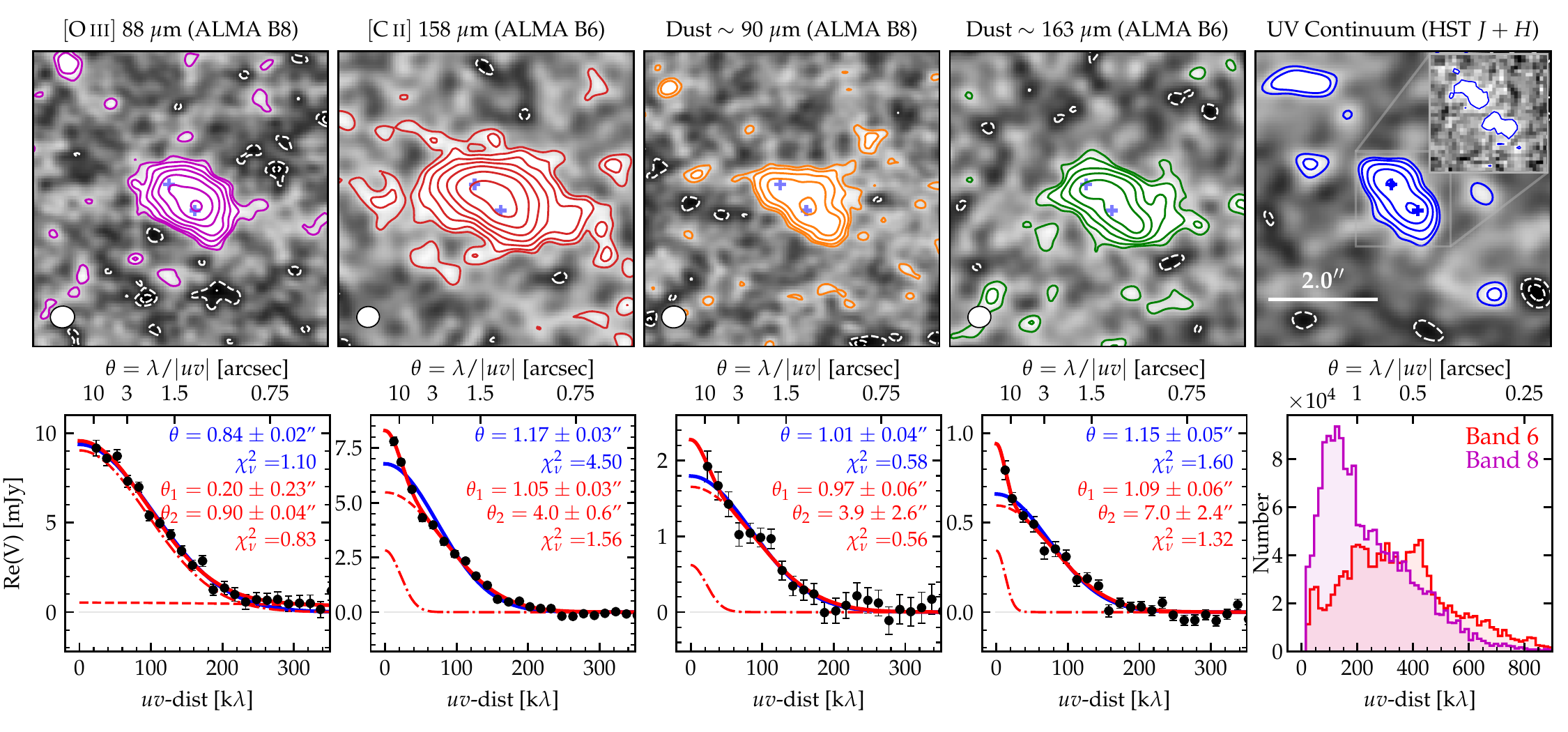}	
	\caption{{\it Top:} Image plane maps of emission from \oiii\ $88~\mu$m, \cii\ $158~\mu$m, band 8 dust continuum, band 6 dust continuum, and the rest-frame UV continuum in \zD. 
	In each panel, white dashed contours show the $-3\sigma$ and $-2\sigma$ levels, while colored contours show the $2\sigma$ and $3\sigma$ levels and then increase following the Fibonacci sequence.
	A slight {\it uv}-taper of $0.3''$ is applied to both band 6 maps to match the resolution of the band 8 maps, and the ellipses on the bottom left indicate the beam size. 
	The rest-frame UV continuum is measured by stacking the {\it HST} $H$ and $J$-bands, masking out nearby sources, and convolving with a Gaussian kernel constructed to match the ALMA beam in band 8.  
	For clarity, we show the unconvolved (high resolution) {\it HST} map in the inset panel for the image plane. 
	We pinpoint the two peaks of the UV continuum emission on each panel in the image plane with two blue crosses. 
	All panels show the same $5.6'' \times 5.6''$ area on the sky.
	{\it Bottom:} Visibility profiles of each ALMA observation. The amplitude is extracted from the real part of the complex visibility in bins of {\it uv}-distance, which corresponds inversely to the observed angular scale $\theta$. 
	We fit each visibility profile to a single Gaussian, and find that this is a good fit for the \oiii\ line and dust continuum. 
	For \cii, however, a double Gaussian model provides a better fit, indicating the presence of significant extended emission. The bottom right panel shows histograms of {\it uv}-dist for the two ALMA bands. The detection of the extended \cii\ structure is not due to an increased sensitivity to large-scale structure (i.e., small {\it uv}-dist) in band 6.}\label{fig:maps}
\end{figure*}

\section{Results}\label{sec:results}

\subsection{2D Maps of \cii, \oiii, UV, and FIR emission}
\label{sec:maps}

The top row of Figure~\ref{fig:maps} presents intensity maps of the \oiii\ and \cii\ lines as well as  the dust continuum at $\sim 90~\mu$m and $\sim 163~\mu$m and the UV continuum as probed by stacking the {\it HST} $J$ and $H$ bands.
Blue crosses in the maps indicate the positions of the two UV continuum peaks, one in the southwest and one in the northeast. 
The peaks of \oiii\ and \cii\ maps are likely to be co-spatial with the southwest UV-bright region. 
This is also where the peak of the dust continuum emission lies, suggesting that this southwest component is the more actively star-forming region of the two. 
The consistency of the peak positions of the \oiii\ and \cii\ with the FIR and UV continuum indicates that the majority of the line emission likely arises from the young star-forming regions visible in UV.

These images indicate that the \cii\ emission is likely extended relative to the \oiii, FIR, and UV.
\cii\ emission is detected at $>2\sigma$ out to $r \sim 2$--$3''$ from the galaxy center.
Furthermore, the dust continuum at $90~\mu$m (band 8) appears more compact than at $163~\mu$m (band 6), which we will revisit in Section~\ref{sec:Tdust}. z

\begin{figure}
	\centering
	\includegraphics[width=\linewidth]{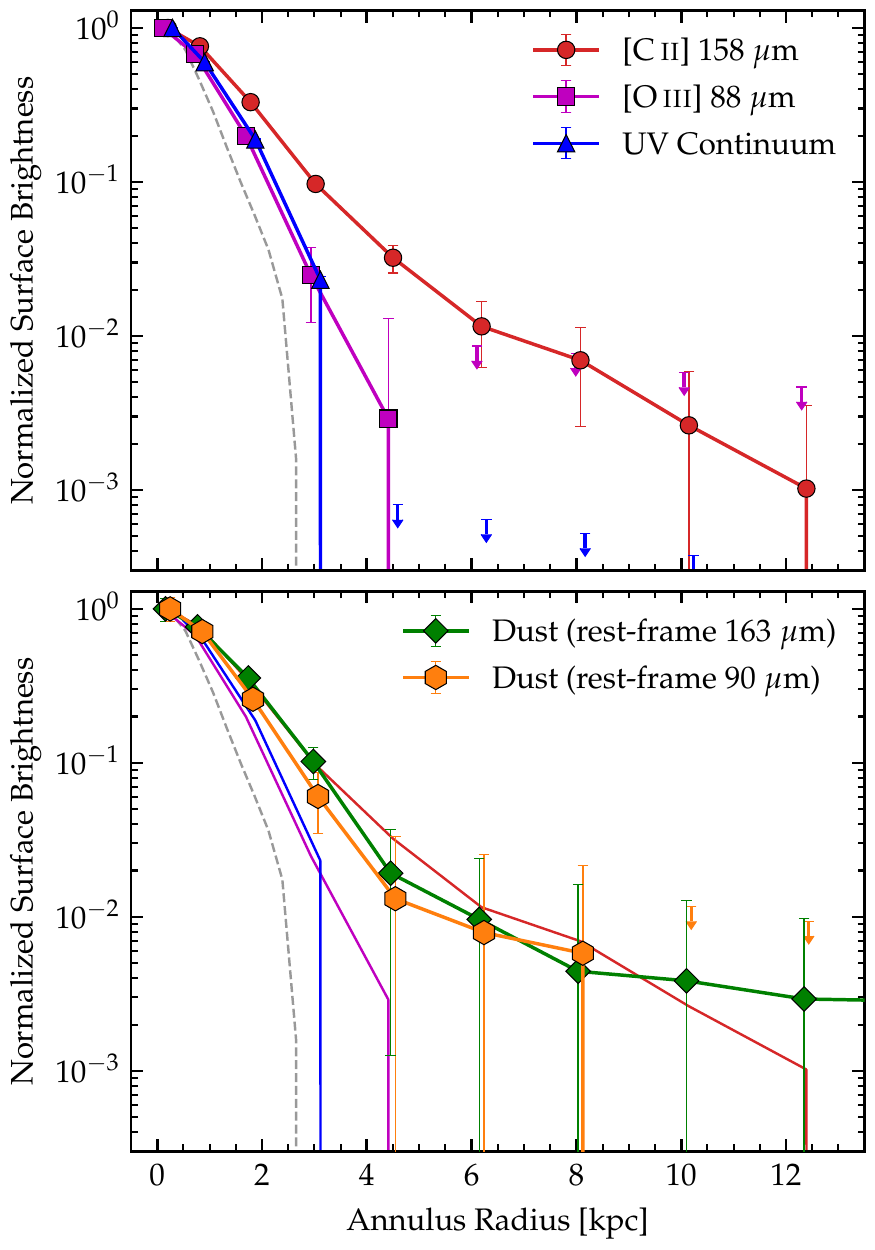}	
	\caption{Radial surface brightness profiles for \zD\ in the source plane. 
	In the top panel, emission from \cii, \oiii, and UV continuum are shown in red, magenta, and blue, as in Figure 1. 
	In the bottom panel, we show emission from the dust continuum at rest-frame $163~\mu$m (green) and $90~\mu$m (orange) over the profiles for \cii, \oiii, and the UV.
	The grey dotted line show the profile of the ALMA PSF.
	Error bars show $1\sigma$ errors on the mean surface brightness in the annulus, computed as the image ${\rm rms}/\sqrt{N}$, where $N$ is the number of full beams in the annulus. 
	When the profile first drops below $0$, we consider it dominated by noise, and thus show $1\sigma$ upper limits.
	We detect extended \cii\ emission out to 12 kpc from the galaxy center, inconsistent with the \oiii\ and UV profiles out to 6--8 kpc.
	The dust continuum follows a similar surface brightness profile to \cii.}\label{fig:profiles}
\end{figure}

\subsection{Spatial Extent of the \cii\ Emission}\label{sec:morphology}

\subsubsection{$uv$-plane Visibility Profiles}\label{sec:vis}

We also analyze the spatial structures of the different emission in the $uv$-visibility plane using the python package \texttt{uvplot} \citep{uvplot}.
In the bottom panel of Figure \ref{fig:maps_uv}, we show the real part of the visibility amplitude in bins of the $uv$-distance, which is inversely proportional to the observed angular scale $\theta$.

We apply a single and double  Gaussian fits to the \oiii, \cii, and dust continuum profiles (shown in blue and red, respectively). 
Based on the reduced chi-squared ($\chi_\nu^2$) statistic, the single Gaussian is a reasonably good model for the \oiii\ line and the dust continuum in both bands, with consistent FWHMs of $\sim 1.0''$; on the other hand, for the \cii\ line, we find a significant offset from the best-fit single Gaussian at the shortest $uv$-distance regime much beyond the error scales. 
The double Gaussian shows a better fit for the \cii\ line, lowering the $\chi_\nu^2$.
For the compact and extended \cii\ components, we derive a best-fit FWHM of $1.05\pm 0.03''$ and $4.0 \pm 0.6''$, respectively.  
The FWHM of the compact \cii\ component is almost the same as those of the \oiii\ line and the dust continuum, while the extended \cii\ component is $\sim 4$ times larger than the compact component. 
These results suggest that there exists an independent extended component surrounding the central galaxy, which is reminiscent of the Ly$\alpha$ halo universally identified around normal star-forming galaxies at $z\sim 3$--$6$ \citep[e.g.,][]{momose_diffuse_2014,leclercq_muse_2017}.
The size ratio of the extended and compact components is comparable to the previous ALMA deep \cii\ stacking results of $\sim 5$--$6$ estimated from galaxies and quasars at $z\sim6$ \citep{fujimoto_first_2019,novak_no_2020}. 

We note that the existence of the extended \cii\ emission, despite no extended component in the \oiii\ emission, is likely not just a result of the data depth or maximum recoverable scale. 
The S/N at the peak pixel in the \cii\ intensity map is $42.9$, almost comparable to \oiii\ ($37.7$), only different by $\sim 10\%$. 
Moreover, as the histogram of {\it uv}-dist in Figure~\ref{fig:maps} indicates, the band 8 observations cover more short baselines (larger angular scales) than band 6, due to the different antenna configurations (C-4 and C-3 for bands 6 and 8, respectively).

Despite the larger error bars, we note that the profiles of the dust continuum in both bands show some excess emission at large angular scale (low $uv$-dist), similar to the \cii\ line. 
Although the lower S/N of this excess trend prevents us from giving a definitive conclusion for the existence of the extended dust component, we examine the potential extended dust structure also in the image-based analysis (Section \ref{sec:radial}).

\subsubsection{Image-based Radial Profiles
}
\label{sec:radial}

We derive radial surface brightness profiles, based on images for both emission lines and the UV and FIR continuum.
We conduct this analysis in the source plane, as described in Section~\ref{sec:spr}. 
For the radial profiles, we use the midpoint between the two UV-bright peaks as the center.

Figure~\ref{fig:profiles} shows radial profiles in the source plane for \cii, \oiii, and UV continuum (top) and for the dust continuum at $90~\mu$m and $163~\mu$m (bottom), normalized at their peaks. 
Points show the mean surface brightness in circular annuli whose width increases at increasing distances, from $~0.4$ kpc to $1$ kpc. 
Error bars show the standard error of the mean. 
For ease of readability, we show these radial profiles out to the first point at which the mean surface brightness drops below 0.
At this point and beyond, we consider the profile dominated by noise fluctuation and instead show only $1\sigma$ upper limits.

Both \oiii\ and UV continuum emission are compact (though extended relative to the PSF) dropping to below 0 at radii of $\sim 4.5$ and $\sim 3$ kpc, respectively. 
In contrast, \cii\ emission is detected out to $r \sim 12$ kpc. 
Furthermore, the \cii\ profile shape is inconsistent with the \oiii\ profile out to $\sim 6$ kpc, and inconsistent with the UV profile out to $\sim 8$ kpc.

Dust continuum emission appears to be extended with a similar radial profile as the \cii\ line.
This is true at both $163~\mu$m and $90~\mu$m, though the latter wavelength drops to 0 at $\sim 8$ kpc while the former extends out to $\sim 12$ kpc.\footnote{We note that drop off in the dust continuum profile at $90~\mu$m is not due to a lack of sensitivity. The peak S/N is about the same for the band 6 and band 8 dust continuum maps ($\sim 26$ vs.~$\sim 23$, respectively). Moreover, our band 8 dataset is more sensitive to large-scale structure as it samples more short baselines (Fig.~\ref{fig:maps}).}
Both continuum profiles are inconsistent with UV and \oiii\ at radii of $r \sim 2$--$4$ kpc, more in line with the \cii\ profile. 
As with the visibility-based profiles, the large error bars at larger scales prevents robust evidence of extended emission. 
Nevertheless, the consistent positive surface brightness in the dust continuum may suggest some real extended emission. 
The presence of such extended dust continuum emission would be in contrast to previous ALMA results for star-forming galaxies at $z\sim 4$--$7$, and we discuss this further in Section~\ref{sec:discussion}.

\begin{figure}
	\includegraphics[width=\linewidth]{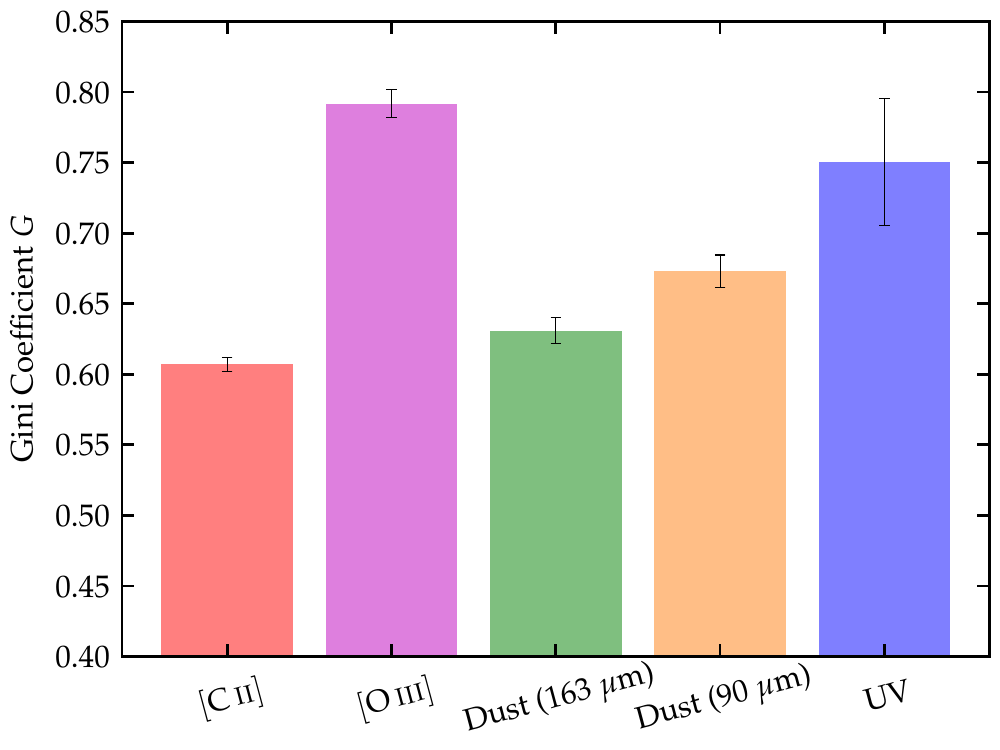}
	\caption{Gini coefficients for \zD. We compute the Gini coefficient for each map following Equation~\ref{eq:gini} and using a circular aperture of radius $2''$. 
	Error bars show $1\sigma$ uncertainty on the Gini coefficient measured in a Monte-Carlo fashion; specifically, for $n=1000$ iterations, we sum the underlying image with a randomized noise map (corresponding to the image rms and convolved with the ALMA beam), and compute the resulting Gini coefficient. 
	Error on the Gini coefficient is inversely correlated with the S/N of the map, but all maps have sufficient S/N to constrain the Gini coefficient within $\sim 5\%$.
	While the \oiii\ line and UV continuum are very concentrated, the \cii\ line FIR continuum are more uniform.}\label{fig:gini}
\end{figure}

\begin{figure*}
    \centering
    \includegraphics[width=0.9\linewidth]{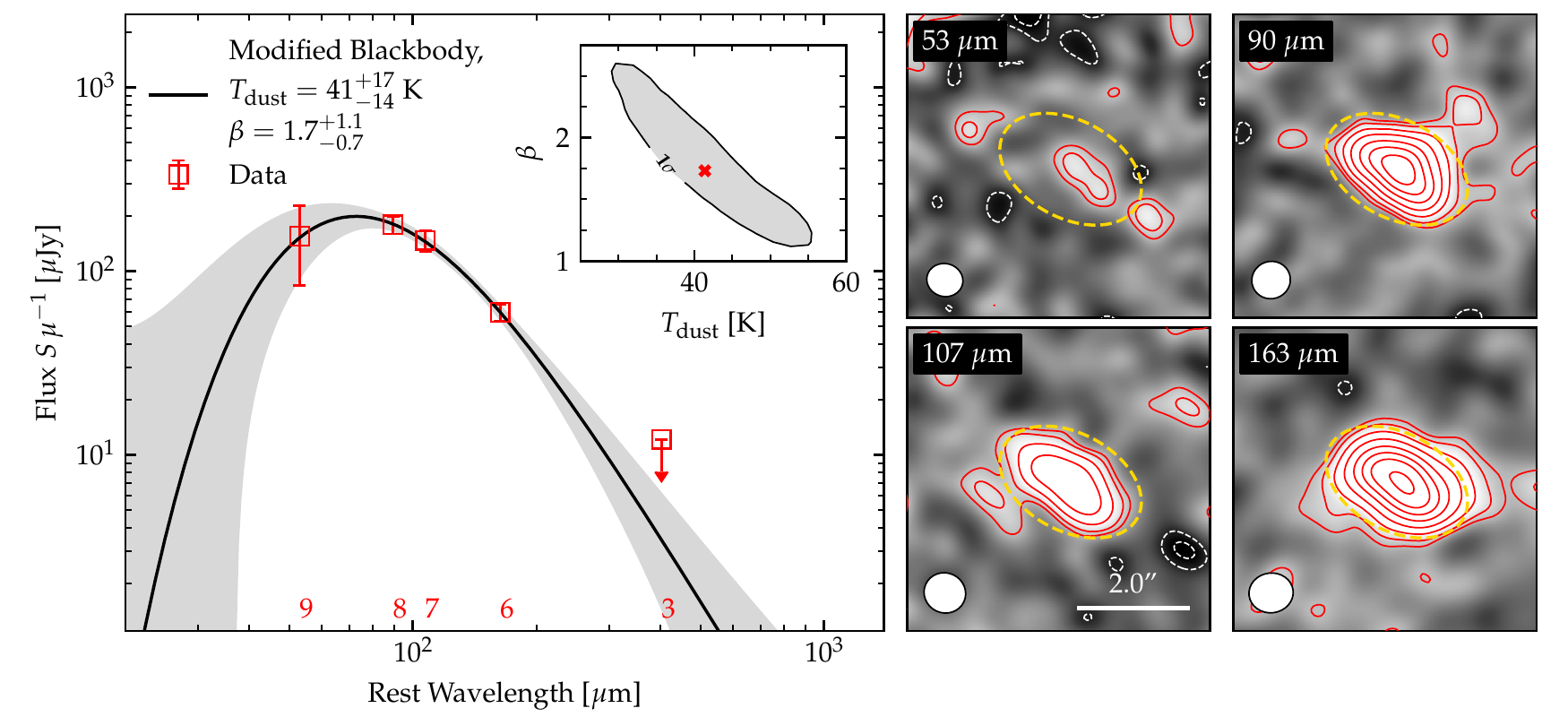}
    \caption{The far-infrared SED of A1689-zD1. We fit a modified blackbody spectrum to ALMA photometry from bands 9, 8, 7, and 6, derived from $2.8'' \times 1.8''$ elliptical apertures placed over the continuum maps shown in the right panels. We additionally constrain the SED fitting according to the 3$\sigma$ upper-limit from the non-detection of continuum in band 3 (rest-frame $\sim 400~\mu$m). The grey shaded region shows the 68\% range of best-fit SEDs determined from varying the flux values according to their uncertainties in a Monte-Carlo fashion. The inset panel plots the $1\sigma$ contour of the probability distribution of $T_{\rm dust}$ and $\beta$ values derived from the Monte-Carlo method, and highlights the degeneracy between the two parameters. 
    We find a best-fit dust temperature of $T_{\rm dust} = 41_{-14}^{+17}$ K and dust emissivity index of $\beta = 1.7_{-0.7}^{+1.1}$, where the error represents the $68\%$ confidence interval. The images on the right show the four continuum detections, with contours denoting the $2,3,5,8,12,16,20,25 \times \sigma$ levels and dashed yellow lines denoting the aperture.}
    \label{fig:SED}
\end{figure*}

\subsubsection{Gini Coefficients}\label{sec:gini}

While radial profiles give a quantitative sense as to the extent of \cii\ emission, they may also be sensitive to the ``clumpiness'' of the emission. 
Recent analysis has shown that \zD\ is a highly clumpy object, requiring multiple spatial/spectral components to fit its \cii\ and \oiii\ datacubes \citep[][Knudsen et al. in prep]{wong_alma_2022}. 
As such, the precise choice of center may impact the shape of the surface brightness profile. 
In the imaging analysis presented thus far, we adopt the midpoint between the two UV-bright peaks as the center; however, in order to test against this possible bias, we additionally compute the Gini coefficient for each emission.
The Gini coefficient ($G$) is a metric borrowed from econometrics, and was originally created to summarize the degree of income inequality in a population. 
It was first used to assess galaxy morphology by \citet{abraham_new_2003}, who showed that it can, to first order, be interpreted similarly to other morphological indicators such as concentration. 
The advantage of using $G$ in this work is that it is non-parametric: it does not depend on the galaxy having a well-defined center, but rather only indicates whether the bulk of the total intensity is contained within a few pixels or spread evenly across many. 
A high value of $G$ ($\sim 0.75$) indicates that the emission from the object is concentrated in one or more bright nuclei, while a low value of $G$ ($\sim 0.25$) indicates that the emission is more uniform. 

We compute the Gini coefficient following \citet{lotz_nonparameteric_2004} by first sorting all $n$ pixel values $X_i$ into increasing order and then computing
\begin{equation}\label{eq:gini}
    G = \frac{1}{\bar{X} n(n-1)}\sum_{i}^n (2i-n-1)X_i
\end{equation}
We compute the error on $G$ in a Monte-Carlo fashion by varying the underlying map by a randomized ``noise map,'' convolved with the ALMA beam.
Since the Gini coefficient it sensitive to the number of pixels included, we compute $G$ based on all pixels within a circular aperture of radius $r = 2''$ for each map. 
Moreover, while the Gini coefficient can be sensitive to the S/N of the map, this effect is negligible above a per-pixel S/N of $\sim 10$ \citep{lisker_gini_2008}, and the sensitivity to the S/N is accounted for by the Monte-Carlo error computation.

Figure~\ref{fig:gini} shows the Gini coefficient computed in the source plane for each map. 
Our Gini coefficient results reaffirm the conclusions from the radial profiles: \oiii\ and UV emission are compact and concentrated in 1 or 2 bright nuclei, while \cii\ and dust continuum emission are spread more evenly. 
Interestingly, the Gini coefficient for the dust continuum suggests a difference between the different wavelengths: the dust continuum at shorter wavelengths is more concentrated than at higher wavelengths.
This may suggest that the dust temperature may vary in space, with hotter dust (which peaks at shorter wavelengths) more concentrated than colder dust (which peaks at higher wavelengths).
The spatially-resolved dust temperature distribution is addressed further in Section~\ref{sec:Tdust}.

\subsection{Spatially-Resolved Dust Temperature}\label{sec:Tdust}

The rich FIR continuum data for \zD\ facilitates accurate determinations of the dust temperature. 
Importantly, band 9 (rest-frame 53~$\mu$m) data is taken, which helps significantly to constrain the FIR SED at shorter wavelengths.
However, the low S/N of the band 9 observations prohibits using this lower wavelength data to constrain the dust temperature in a spatially-resolved sense.
We opt instead to use the ratio between bands 6 ($163~\mu$m) and 8 ($90~\mu$m) to spatially-resolve the dust temperature; however, this method depends on the assumed dust emissivity index, $\beta$, which usually ranges from $1$--$2.5$ \citep{casey_farinfrared_2012}. 
Therefore, we first determine the best-fit $\beta$ value via a modified blackbody fit on the galaxy as a whole, using continuum data from bands 3, 6, 7, 8, and 9. 
We then derive the spatially-resolved dust temperature map, assuming that $\beta$ remains constant in space.

Figure~\ref{fig:SED} shows the FIR SED of A1689-zD1. 
Measurements of the dust continuum flux from bands 6 ($163~\mu$m), 7 ($107~\mu$m), 8 ($90~\mu$m), and 9 ($53~\mu$m) are shown, and a $3\sigma$ upper limit is shown for the continuum in band 3 ($\sim 400~\mu$m), which is not detected. 
All photometry is computed in the $2.8'' \times 1.8''$ elliptical aperture shown over the dust continuum images on the right.
This aperture is adopted consistently across all bands and is determined by eye to correspond roughly to the \oiii-emitting region.
The aperture is large enough to cover the bulk of the dust continuum in bands 6--8, but not so large as to include outskirt noise fluctuation in band 9 and significantly reduce the S/N in that band. 
Since band 9 holds most of the constraining power in our MBB fit, it is critical to focus on this central region where the band 9 continuum is robustly detected.
Error bars in Figure~\ref{fig:SED} show the combined uncertainty, including both statistical uncertainty measured from the image rms, and an additional systematic uncertainty of $10\%$ ($20\%$ in band 9) arising from flux calibration, as specified in the ALMA Cycle 8 proposer's guide.\footnote{See \S A.9.2, \url{https://almascience.nrao.edu/documents-and-tools/cycle8/alma-proposers-guide}}
We fit a modified blackbody spectrum to the data points, correcting for the decreasing temperature contrast between the target and background due to the heating of dust by the CMB following \citet{dacunha_effect_2013}. 
We perform this fitting using a Monte-Carlo approach, allowing $T_{\rm dust}$ and $\beta$ to vary freely with flat priors. 
We determine a best-fit dust temperature of $T_{\rm dust} = 41^{+17}_{-14}$ K and dust emissivity index $\beta = 1.7^{+1.1}_{-0.7}$. 
The inset panel shows the distribution of $T_{\rm dust}$ and $\beta$ values derived from the MCMC fitting, and highlights the inherent degeneracy between the two parameters. 
The larger uncertainty in our $T_{\rm dust}$ and $\beta$ measurements compared to \citet{bakx_accurate_2021} is due to our use of a 20\% uncertainty on the band 9 measurement.

\begin{figure}
	\centering
	\includegraphics[width=\linewidth]{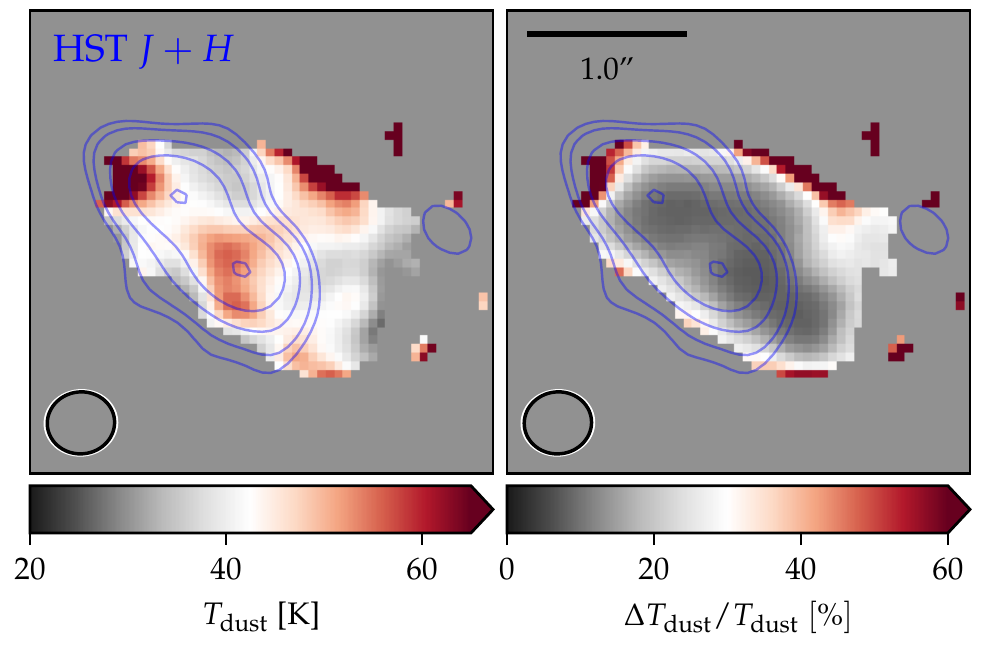}
	\caption{Dust temperature map for \zD. Blue contours show the UV continuum as in Figure~\ref{fig:maps}. The right panel shows the fractional uncertainty on the inferred dust temperature. We see high $T_{\rm dust}$ in the southeast UV-bright region, corresponding to the bright \cii\ and \oiii-emitting region, but the two other high $T_{\rm dust}$ regions are likely a product of noise.}\label{fig:Tdust}
\end{figure}

To extend the analysis in a spatially-resolved fashion, we then fix $\beta = 1.7$ as a fiducial value. 
We note that, with variations in the dust chemical composition and crystallinity, it would be possible for $\beta$ to vary in space with $T_{\rm dust}$ fixed. 
However, we have fit the full FIR SED in two spatially-resolved apertures and found $< 1\sigma$ difference in $\beta$ between the two. 
Therefore, we consider it reasonable to fix $\beta$ and explore the resulting variation in $T_{\rm dust}$. 
We do so by deriving a relationship between the observed rest-frame $90~\mu$m to $163~\mu$m continuum ratio (corrected for CMB effects) and the dust temperature. 
Figure~\ref{fig:Tdust} shows the spatially resolved dust temperature map for pixels where the dust continuum is detected (at $2\sigma$) at both $163~\mu$m and $90~\mu$m. 
We show in the right panel the fractional uncertainty on the inferred dust temperature.
While we see three high dust temperature regions, the two on the far northeast and northwest sides of the galaxy are likely a product of noise fluctuation, as the uncertainty is $> 60\%$.

We overplot contours showing the stacked {\it HST} $J+H$-band image. 
The southeast UV-bright region corresponds to a peak in the dust temperature of $\sim 50$ K, while the northwest region corresponds primarily to a noisy peak. 
However, the high dust temperature in the galaxy center stretches out towards both noisy peaks, with low errors. 
Similarly, the far southwest of the map shows a  higher temperature than the minimum, with low errors.
These regions likely correspond to distinct clumps in the galaxy ISM, which have been identified in the galaxy spectrum \citep[][Knudsen et al. in prep]{wong_alma_2022}.

\begin{figure}
	\includegraphics[width=\linewidth]{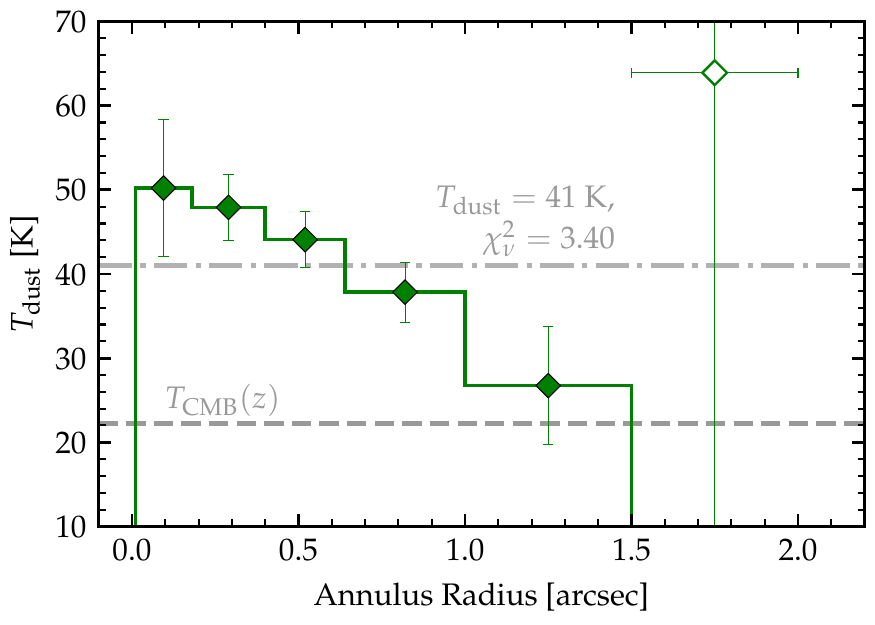}	
	\caption{Radial profile of the dust temperature in \zD. 
	We derive the dust temperature from the ratio of the continuum at $90~\mu$m to the continuum at $163~\mu$m, correcting for CMB effects. 
	As in Section~\ref{sec:morphology}, we use {\it uv}-tapered maps. 
	Error bars show $1\sigma$ uncertainty on the inferred dust temperature derived in a Monte-Carlo fashion from the uncertainty on the continuum ratio.
	We show the furthest bin as an open diamond, as the S/N is too low to provide a stringent constraint on the dust temperature. 
	Despite this uncertainty, the dust temperature appears to decrease to larger radii, approaching the CMB temperature.}\label{fig:Tdust_profile}
\end{figure}

In addition to estimating $T_{\rm dust}$ on a pixel-by-pixel basis, we also endeavor to estimate the spatially-resolved $T_{\rm dust}$ on kpc scales. 
Figure~\ref{fig:Tdust_profile} shows the inferred dust temperature as a function of distance, in arcsec, based on image-plane radial profiles of the continuum maps at both wavelengths. 
We use circular annuli of variable width out to $2''$ from the galaxy center, which corresponds to $\sim 7$ kpc when accounting for the gravitational lensing. 
We use image-plane maps for this analysis to minimize any additional uncertainty brought about by the source-plane reconstruction.
Error bars show $1\sigma$ uncertainty on the dust temperature. 
The radial profile suggests that the dust temperature decreases with increasing distance from the galaxy center, starting at $\sim 50$ K within $0.25$'' and dropping to $\sim 30$ K beyond $1$''. 
At further distances, it appears to increase again, though our signal-to-noise becomes too low to robustly measure the dust temperature.

We explore the feasibility of a $T_{\rm dust}$ gradient by applying a constant model in which the dust temperature is fixed at the best-fit spatially-integrated value ($T_{\rm dust} = 41$ K) everywhere. 
This model results in a reduced chi-squared of $\chi_\nu^2 = 3.4$. 
In contrast, a simple linear model for the decreasing trend agrees within the error in every annulus.
These results indicate that a model with constant $T_{\rm dust}$ is unlikely to be consistent with the data. 
While the errors in our results are large, this analysis suggests that deep ALMA observations, with the aid of gravitational lensing, may be able to observe the $T_{\rm dust}$ gradient at $z>7$.

\begin{figure*}
	\centering
	\includegraphics[width=\linewidth]{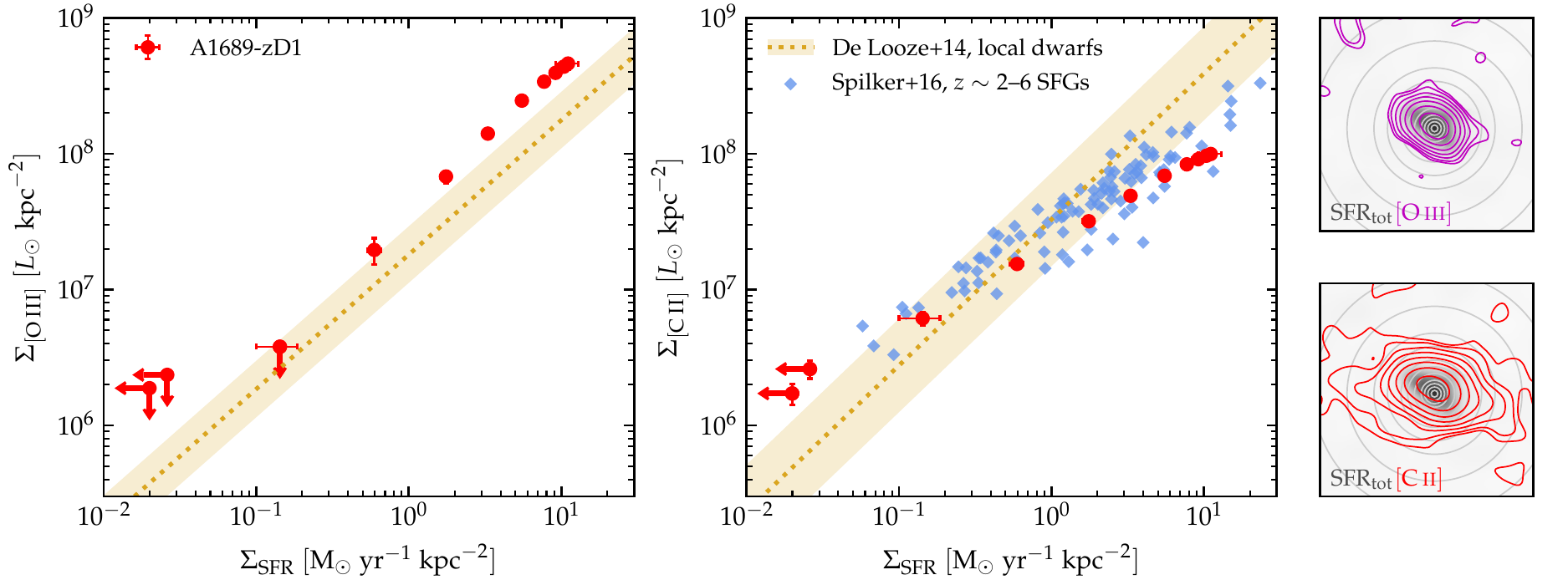}
	\caption{The spatially-resolved \Sigoiii-\Sigsfr\ and \Sigcii-\Sigsfr\ relations for \zD. 
	Red points show spatially-resolved data derived from circular annuli of varying widths.
	Gold lines show empirical spatially-resolved, local universe relation from \citet{delooze_applicability_2014}, while blue points show measurements of individual $z \sim 2$--$6$ star-forming galaxies from \citet{spilker_alma_2016}, assuming $\Sigma_{\rm SFR} = 1.48\times 10^{-10}\Sigma_{\rm FIR}$.
	The right panels show the \oiii\ and \cii\ contours plotted over the SFR map, with the borders of the circular annuli.}\label{fig:SigOIII_SigCII_SigSFR}
\end{figure*}

\subsection{The \Sigcii-\Sigsfr\ and \Sigoiii-\Sigsfr\ Relations}\label{sec:SFR}

Spatially-resolving the dust temperature facilitates robust estimates of the spatially-resolved total IR luminosity ($8$--$1000~\mu$m) and obscured SFR. 
For each pixel, we compute the total IR luminosity by integrating a modified blackbody with $\beta = 1.7$ and $T_{\rm dust}$ set to the inferred dust temperature shown in Figure~\ref{fig:Tdust}, or, if the temperature was not well determined,\footnote{We consider the temperature undetermined if either one or both of band 6 or 8 continuum maps were not detected a S/N $> 2$, or if the resulting fractional uncertainty on the dust temperature exceeds $30\%$.} with $T_{\rm dust} = 30$ K. 
We then compute the obscured SFR following the \citet{murphy_calibrating_2011} calibration as
\begin{equation}
  {\rm SFR}_{\rm IR}~[M_\odot/{\rm yr}] = 3.88\times10^{-44} \times L_{\rm IR}\,\mu^{-1}~[{\rm erg}/{\rm s}]
\end{equation}
We additionally compute the unobscured (UV-based) SFR following the \citet{murphy_calibrating_2011} calibration as
\begin{equation}
   {\rm SFR}_{\rm UV}~[M_\odot/{\rm yr}] = 4.42\times10^{-44} \times L_{\rm UV}\,\mu^{-1}~[{\rm erg}/{\rm s}]
\end{equation}
where $L_{\rm UV}$ is the UV luminosity measured from the stacked {\it HST} $J$+$H$ band image. 
This probes the rest-frame wavelength range $1350$~\AA $< \lambda < 2050$~\AA, comparable to the GALEX FUV transmission curve used in the calibration of \citet{murphy_calibrating_2011}. 
The calibrations from \citet{murphy_calibrating_2011} assume a \citet{kroupa_variation_2001} IMF and a constant star-formation history with an age of $\sim 100$ Myr.
Uncertainties on the derived SFRs include the uncertainties on the ALMA/{\it HST} maps as well as the uncertainty in the dust temperature (for SFR$_{\rm IR}$). 
Summing over the elliptical aperture used for photometry in Section~\ref{sec:Tdust}, and assuming $\beta=1.7$, we derive an obscured SFR of ${\rm SFR}_{\rm IR} = 32 \pm 1~M_\odot~{\rm yr}^{-1}$ and an unobscured SFR of ${\rm SFR}_{\rm UV} = 4.95 \pm 0.03~M_\odot~{\rm yr}^{-1}$, resulting in a total SFR of $\sim 37 \pm 1~M_\odot~{\rm yr}^{-1}$ and an obscured fraction of $\sim 87\%$. 
The high obscured SFR is inline with recent estimates for \zD\ from \citet{bakx_accurate_2021}.
We derive a total infrared luminosity of $L_{\rm IR}\,\mu^{-1} = (2.16 \pm 0.07)\times 10^{11}~{\rm L}_\odot$.

To compare the derived SFR to \cii\ and \oiii\ emission, we place circular annuli of varying width across the tapered maps, and compute the SFR and line surface densities in each annulus.
Figure~\ref{fig:SigOIII_SigCII_SigSFR} shows the spatially-resolved \Sigoiii-\Sigsfr\ and \Sigcii-\Sigsfr\ relations with the empirical, local universe relation from \citet{delooze_applicability_2014} and (for \Sigcii) the results for $z \sim 2$--$6$ dusty star-forming galaxies from \citet{spilker_alma_2016}. 
We see that the \oiii\ line roughly traces the SFR following the relation for local, low-$Z$, dwarf galaxies, with a slight ($\sim 0.3$ dex) excess in regions with high \Sigsfr\ (i.e., the galaxy center). 
By contrast, we see a $\sim 0.7$ dex deficit of \cii\ in the galaxy center, relative to local dwarfs.
This deficit is consistent with the \cii-FIR deficit in high redshift DSFGs observed by \citet{spilker_alma_2016}, which has also been observed in local IR luminous galaxies \citep{diaz-santos_extended_2014}. 
The deficit at high \Sigsfr\ diminishes at lower \Sigsfr, with the two largest annuli showing \cii\ detections despite non-detections of UV or IR continua. 
The \oiii\ line is also not detected in these annuli.

\subsection{Central \cii\ and \oiii\ Line Profiles}
\label{sec:line_profiles}

\begin{figure*}
	\centering
	\includegraphics[width=0.8\linewidth]{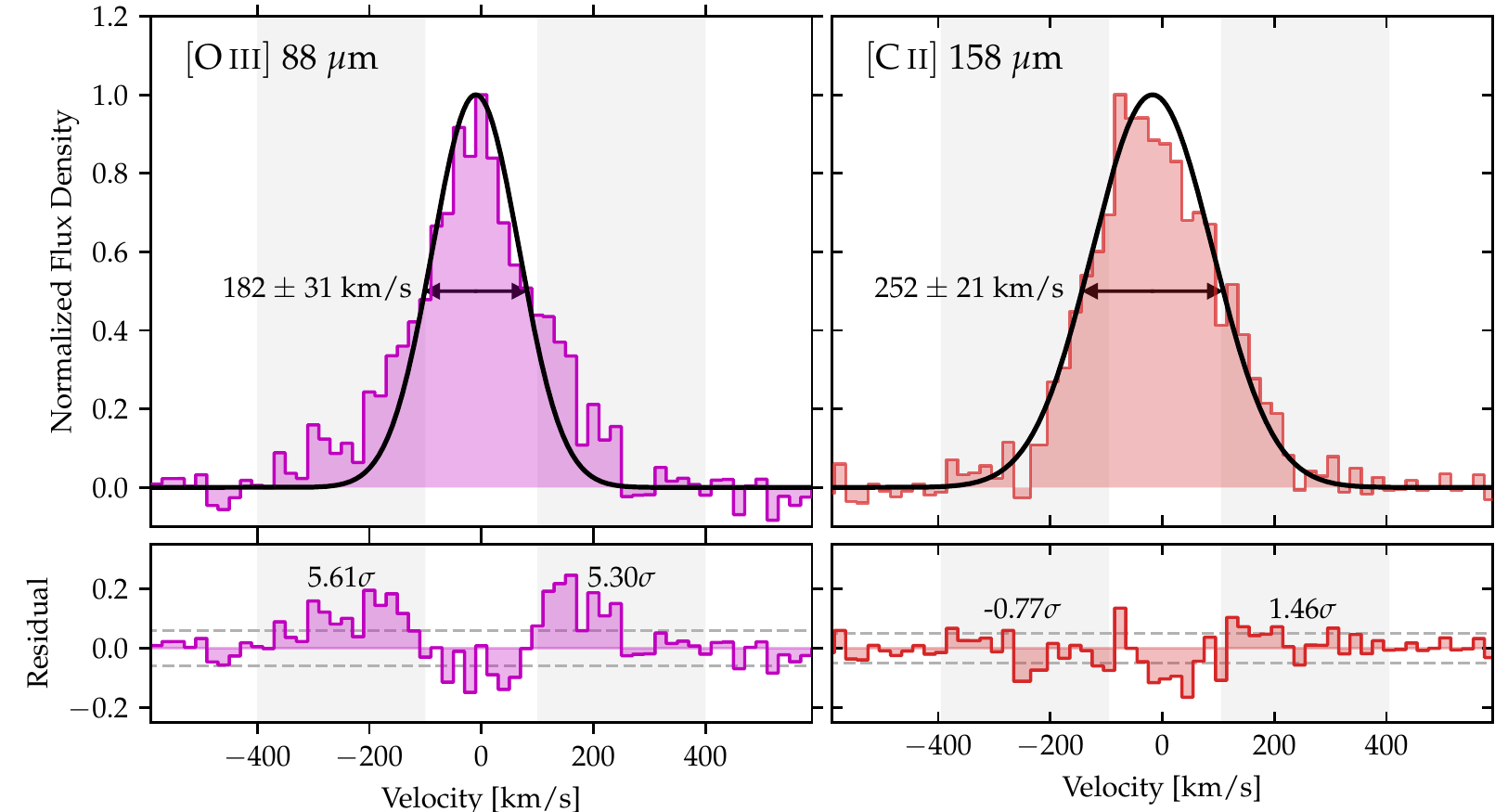}
	\caption{Line profiles for \oiii\ (left) and \cii\ (right) in a $0.25''$ aperture centered on the high $\Sigma_{\rm SFR}$ region. We apply Gaussian fits to both lines with the FWHM fixed at the measured value. The bottom panel shows the residuals from these fits, annotated with the signal to noise of the residuals integrated across the high-velocity regions at $\pm100$--$400$ km~s$^{-1}$. While the \cii\ line is broader than the \oiii\ line, the \oiii\ line shows significant high-velocity residuals, suggesting the possibility of ongoing hot ionized outflows from the galactic core.}\label{fig:line_profiles}
\end{figure*}

We also investigate the \cii\ and \oiii\ line profiles in a spatially-resolved manner. 
In these emission line spectra, high-velocity outflows in different gas phases may be identified via the ``broad-wings'' that extend past the extent of the best-fit Gaussian profile \citep[see e.g.][]{veilleux_cool_2020}. 
We produce line profiles from the \cii\ and \oiii\ cubes in a small ($0.25''$ radius) aperture centered at the the peak of the $\Sigma_{\rm SFR}$ map, i.e. over the UV$+$FIR-bright region. 
This small radius is roughly the beam size of the fiducial ALMA maps. 
We focus on this region for two reasons: 1) the clumpy nature of \zD\ confuses identification of the broad-wing feature with distinct spectral components, and 2) in a SF-driven outflow scenario, the faint broad-wing feature would be most easily detected in regions of highest $\Sigma_{\rm SFR}$ \citep{davies_kiloparsec_2019}. 
We have verified that this aperture is small enough to probe only one of the five kinematic component components identified in Knudsen et al. (in prep). 
Moreover, as the same aperture is applied to both the \cii\ and \oiii\ maps, any differences in the line profiles corresponds to different kinematics between \cii\ and \oiii\ rather than contamination from different components, which would be seen in both emission lines.

Figure~\ref{fig:line_profiles} shows the line profiles for \cii\ and \oiii\ in this small aperture, with Gaussian fits applied. 
Before performing any Gaussian fits, we determine the FWHM of each line and examine the differences in the line widths.\footnote{We determine the FWHM via linear interpolation between the data above and below $50\%$. We determine the error on the FWHM in a Monte-Carlo fashion, by varying each data point in the line spectrum according to the uncertainty.} 
The \cii\ line width of $252 \pm 21$ km~s$^{-1}$ is significantly larger than the \oiii\ line width of $182 \pm 31$ km~s$^{-1}$. 
The smaller line width for the \oiii\ line is consistent with hydrodynamic simulations \citep[e.g.][]{katz_probing_2019b,katz_nature_2022}, as the \oiii\ tends to trace younger star-forming regions which tend to have a lower velocity dispersion than the system as a whole. 
The larger line width for \cii\ may also be related to the spatial extent of the \cii\ emission. 
If the extended emission were spherically symmetric, then the larger volume of gas observed may capture more complicated kinematics along the line of sight and thus contribute to the larger line width.

To test for the outflow signature, we fit a Gaussian to each line with the FWHM fixed at the measured value. 
We see that the \cii\ line is well fit by a single Gaussian, with a reduced chi-squared of $\chi_\nu^2 = 1.18$. 
By contrast, the \oiii\ line is not well fit by a single Gaussian, with a reduce chi-squared of $\chi_\nu^2 = 2.16$ ($p<10^{-7}$). 
This owes to significant excess flux in high-velocity regions, despite the narrow line width overall. 
We show in the bottom panels of Figure~\ref{fig:line_profiles} the residuals on each Gaussian fit. 
For \oiii, we find significant ($> 5\sigma$) residuals when integrating across the high-velocity tails from $\pm 100$--$400$ km~s$^{-1}$. 
As \cii\ is well-fit by a single Gaussian, we find no significant residuals in these high-velocity tails.

\section{Discussion}\label{sec:discussion}

We have shown in Section~\ref{sec:results} that the \cii\ emission (and possibly also the dust continuum emission) in \zD\ is extended on $\sim 12$ kpc scales relative to \oiii\ and UV emission. 
The \cii\ line emission in $uv$ space is well-fit by a double Gaussian model, with a extended/compact component ratio of $\sim 4$. 
This \cii\ emission deviates slightly from the local \Sigcii-\Sigsfr\ relation, with a ``deficit'' of \cii\ in the galaxy center and a slight excess in the furthest apertures. 
We additionally detect significant large-velocity residuals in the \oiii\ line spectrum at the galactic core, indicating the possible presence of outflow dominated by hot, ionized gas. 
We now discuss in depth the physical interpretation of these observations;
in particular, 1) the description of the extended \cii\ emission as a part of the central galaxy versus a ``\cii\ halo,'' separate from a single galaxy disk and reminiscent of the Ly$\alpha$ halo, and 2) the physical origin of the extended emission.

\subsection{\cii\ halo vs.~Extended disk}\label{sec:halo_vs_disk}

The results presented thus far are reminiscent of the ``Lyman-$\alpha$ halo'' universally identified around normal star-forming galaxies at $z\sim 3$--$6$ \citep[e.g.,][]{momose_diffuse_2014,leclercq_muse_2017}.
However, another possibility is that we are observing the outskirts of the central galaxy disk even out to $r\sim 10$ kpc, and that star-formation in this disk is the source of the extended \cii\ emission \citep[e.g.,][]{novak_no_2020}.
Due to the detection of extended FIR emission (Fig.~\ref{fig:profiles}), we cannot immediately rule out the latter scenario.

While the rest-UV and \oiii\ sizes are similarly compact, we see potential evidence of extended dust emission in both \textit{uv}- and image-based analyses (e.g., Figures~\ref{fig:maps} and \ref{fig:profiles}). 
The presence of extended dust continuum emission would be in contrast to previous ALMA results for star-forming galaxies at $z\sim 4$--$7$ \citep{fujimoto_first_2019, ginolfi_alpinealma_2020, herrera-camus_kiloparsec_2021a}, and would, at first glance, suggest significant obscured star-formation in the outskirts of the galaxy.

One possible explanation for this inconsistency with previous observations is that this extended dust emission is common in the early universe, but too cold and faint to be detected, being embedded in the warm CMB \citep[e.g.][]{dacunha_effect_2013, vallini_cii_2015, zhang_gone_2016, pallottini_zooming_2017, lagache_cii_2018}.
The dust continuum profiles for \zD\ are similar to the UV and \oiii\ line up to $r\sim 3$ kpc, and only diverge at larger distances. 
This faint extended structure may have been missed in previous stacking results, especially if the observed dust temperature gradient is common, as it is in local galaxies \citep{hunt_cool_2015} and as is predicted in cosmological zoom-in simulations for early galaxies \citep[e.g.,][]{behrens_dusty_2018,sommovigo_warm_2020}. 
For this study, the deep integration times in bands 6 and 8, combined with the strong lensing effect (with magnification $\mu \sim 10$) yields an effective sensitivity $\sim 10$ times deeper than previous stacking results \citep[e.g.][]{fujimoto_first_2019, ginolfi_alpinealma_2020}. 
Under this interpretation, many high-$z$ star-forming galaxies may host extended dust continuum emission corresponding to their extended \cii\ emission.
In fact, recent stacking results for $z \gtrsim 6$ massive quasar host galaxies find similar radial profiles for \cii\ and dust \citep{novak_no_2020}, although the physical mechanisms producing extended dust emission might be different between massive quasar host galaxies and sub-L* galaxies. 

It is worth noting that the detection of extended dust continuum emission implies a larger dust mass than what has previously been estimated (e.g.~$M_{\rm dust} = 1.7^{+1.3}_{-0.7}\times 10^7~M_\odot$ from Bakx et al.~\citeyear{bakx_accurate_2021}). 
Indeed, the elliptical aperture applied in Section~\ref{sec:Tdust} fails to capture $\sim 13\%$  of the band 6 continuum flux observed out to $r\sim 1.5''$. 
Based on our spatially-resolved $T_{\rm dust}$ estimates,  
fitting this excess flux to a modified blackbody with $T_{\rm dust} = 30$ K and $\beta = 1.7$ yields a dust mass of $M_{\rm dust} \sim 7\times10^6~M_\odot$. 
This suggests that the dust mass estimate could be increased by $\sim40\%$ compared with the previous estimate of \citet{bakx_accurate_2021} by including the extended dust component. 
If the extended dust component ubiquitously exists around early galaxies, this additional, previously unaccounted for dust mass may have implications for dust evolution in the early universe \citep[e.g.][]{mancini_dust_2015, michalowski_fate_2019, dayal_alma_2022}.

\begin{figure}
    \centering
    \includegraphics[width=\linewidth]{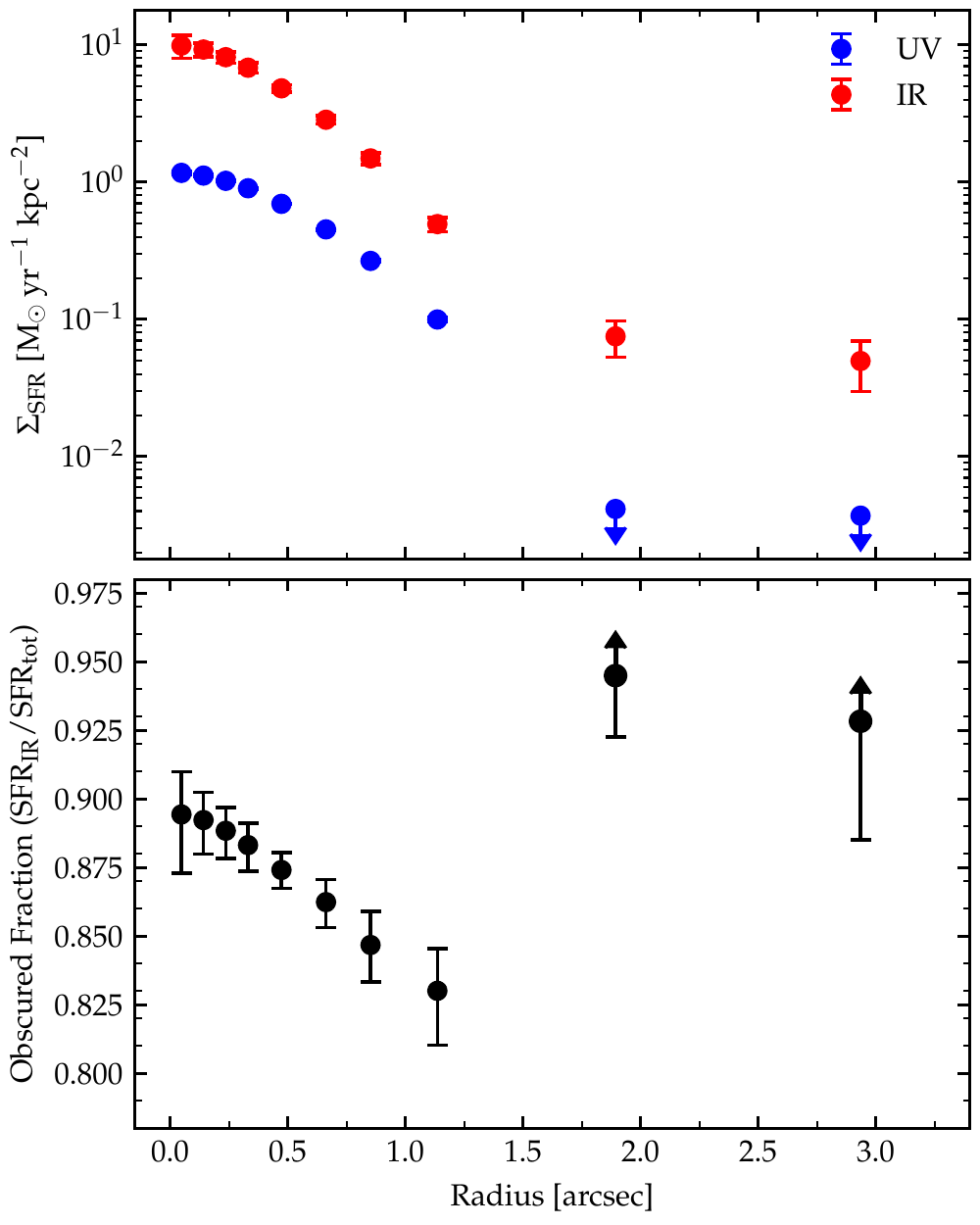}
    \caption{$\Sigma_{\rm SFR}$ and obscured fraction assuming that the extended FIR continuum emission is solely due to localized SF. {\it Top}: Surface density of star formation derived from the UV (blue) and IR (red), as a function of radius in several circular annuli. {\it Bottom}: Obscured fraction of star formation (${\rm SFR}_{\rm IR}/{\rm SFR}_{\rm tot}$) as a function of radius.
    In both panels, arrows indicate $3\sigma$ upper/lower limits.
    Non-detections of the UV SFR in the outer annuli, where IR emission is detected, constrain the obscured fraction to $\gtrsim 93\%$. 
    Such a high obscured fraction in the outskirts is unexpected, and may suggest that the extended FIR emission is not entirely due to SF.}
    \label{fig:obscured_fraction}
\end{figure}

The extended \cii\ and dust continuum emission could be explained by localized star-formation, i.e., an extended SF disk rather than an independent halo component.
Given that the dust temperature at the outskirts of the galaxy approaches the CMB temperature (Figure~\ref{fig:Tdust_profile}), it is possible that we underestimate $\Sigma_{\rm IR}$ in these regions despite correcting for the decreasing contrast following \citet{dacunha_effect_2013}.
However, even in this scenario, the compact rest-UV size of the galaxy implies that the obscured fraction increases up to $\sim 100\%$ towards the outskirts of the galaxy. 
We show in Figure~\ref{fig:obscured_fraction} the UV and IR SFRs (top) and obscured fraction (bottom) as a function of radius in several circular annuli (as in Figure~\ref{fig:SigOIII_SigCII_SigSFR}).
While the obscured fraction decreases with increasing radius up to $r\sim 1''$, the detections of IR SFR despite non-detections of UV SFR yield $3\sigma$ lower limits on the obscured fraction of $\sim 93\%$. 
Such a reversal of the decreasing trend is in contrast to the metallicity gradient expected from simulations, which would continue to decrease to $\sim 1\%$ at the outskirts \citep{pallottini_zooming_2017}.

Therefore, a realistic picture might include some dust heating mechanisms other than localized SF activity in the outskirts of the galaxy, such as heating from the interstellar radiation field (ISRF) or by hot outflows. 
In this scenario, we would need to correct the SFR estimates by removing the extended FIR emission counted as SFR. 
Then, the \Sigcii-\Sigsfr\ relation in the furthest annuli would deviate even more from the local relation, strengthening the interpretation that extended \cii\ emission is powered by other heating mechanisms than localized SF.

At the same time, the \cii\ emission in the outskirts of the galaxy can also be suppressed by the CMB, depending on the gas temperature and density \citep{vallini_cii_2015,kohandel_kinematics_2019}.
Since in general the gas temperature ($\gtrsim 10^2$ K) is much higher than $T_{\rm CMB}$, the detection of the extended \cii\ emission may indicate i) the extended C$^+$ gas maintains a relatively high density to be not significantly affected by the CMB suppression, or ii) the majority of the \cii\ emission in the outskirt regions is actually suppressed and the intrinsic \cii\ luminosity is even greater. 
The latter case would yield even further deviation from the local \Sigcii-\Sigsfr\ relation (right panel of Figure~\ref{fig:SigOIII_SigCII_SigSFR}.

\begin{figure*}
\centering
\includegraphics[width=0.75\linewidth]{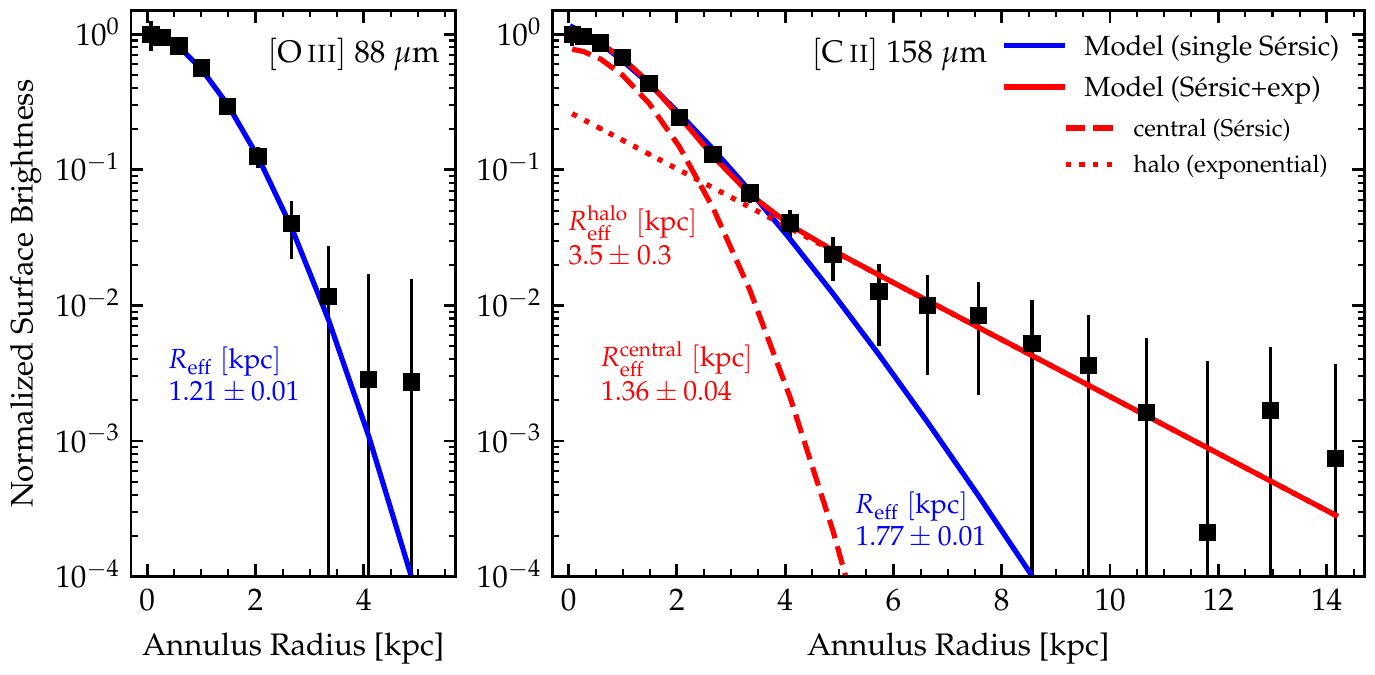}
\caption{Radial profiles of the \oiii\ (left) and \cii\ (right) emission, with two-component Sérsic+exponential profiles applied. The \oiii\ emission is well-fit by a single Sérsic profile with $R_{\rm eff} = 1.21\pm 0.01$ kpc and $n = 0.59\pm 0.02$. By contrast, the \cii\ line requires a central Sérsic profile with $R_{\rm eff}^{\rm central} = 1.36\pm 0.04$ and $n=0.55 \pm 0.03$ as well as an exponential profile with $R_{\rm eff}^{\rm halo} = 3.5\pm 0.3$ kpc. The clear secondary component suggests the existence of a \cii\ halo.}
\label{fig:profile_fitting}
\end{figure*}
 
As \zD\ is a low-mass ($\sim 10^9 M_\odot$), sub-L* galaxy \citep{watson_dusty_2015}, the existence of an extended disk of $r\sim 10$ kpc would be striking. 
The effective radius for such sub-L* galaxies is generally estimated to be $< 0.8$ kpc \citep[e.g.][]{shibuya_morphologies_2015}.\footnote{While the rest-frame UV is not necessarily a perfect tracer of the overall source size, \citet{shibuya_morphologies_2015} find consistent sizes with rest-frame optical observations out to $z\sim 1$. Similarly, \citet{jimenez-andrade_vla_2021} find consistent sizes between UV, optical, and radio wavelengths at $z \sim 3$.}
Following studies of the Lyman-$\alpha$ halo \citep[e.g.][]{hayes_lyman_2014, momose_diffuse_2014, leclercq_muse_2017}, we fit the \cii\ and \oiii\ profiles using a combined Sérsic+exponential model to fit the central and halo components, respectively. 
Figure~\ref{fig:profile_fitting} shows these fits. 
For the \oiii\ line, we find that including a second component is unnecessary, as the emission is well fit by a single Sérsic profile with $R_{\rm eff} = 1.21 \pm 0.01$ kpc and $n = 0.59 \pm 0.02$. 
This is generally consistent with the rest-frame optical and UV sizes measured by \citet{shibuya_morphologies_2015}, though perhaps a bit more extended. 
For the \cii\ line, the single Sérsic model fails to capture the extended emission; instead, we need a central component with $R_{\rm eff}^{\rm central} = 1.36 \pm 0.04$ kpc and $n=0.55\pm 0.03$ and an exponential halo component with $R_{\rm eff}^{\rm halo} = 3.5\pm 0.3$ kpc. 
In addition to the visibility-based analysis (Section~\ref{sec:vis}), the fitting clearly confirms the existence of two separate components, which is reminiscent of the Ly$\alpha$ halo identified around high-$z$ SFGs. 

In summary, the extended dust continuum and \cii\ emission might be explained by the outskirt of an extended galaxy disk. 
But this interpretation suggests an increasing trend of the obscured fraction as a function of radius up to $\sim 100\%$ at large radii. 
Instead, the extended \cii\ structure is more likely explained by an independent gaseous halo, which is evident from both visibility- and image-based profile fitting for the \cii\ emission, and we may call it ``\cii\ halo.'' 
The \cii\ halo and the extended dust emission would have formed via some physical mechanism(s) other than localized SF, which we will discuss in the next section.

\subsection{The Physical Origin of the \cii\ Halo}\label{sec:origin}

Assuming that the extended \cii\ emission represents a distinct \cii\ halo, we now discuss the physical origin of this feature.
\citet{fujimoto_first_2019} present five possible explanations for \cii\ emission on circumgalactic scales: A) star formation from satellite galaxies; B) circumgalactic-scale photodissociation region (CG-PDR); C) circumgalactic-scale \hii\ region (CG-\hii); D) cold streams; and E) outflow. 
For merging galaxies in the center of the protocluster, \citet{ginolfi_alpinealma_2020a} argue a sixth possible explanation: F) turbulence and shocks associated with tidal stripping.

In all cases, some physical process is required to enrich the CGM with carbon and produce the halo of C$^+$ that we observe. 
Assuming that cold streams are composed primarily of pristine gas, only scenarios A), E), and F) include some mechanism to enrich the CGM with carbon. 
Scenarios B) (CG-PDR) or C) (CG-\hii) may certainly ionize the carbon and power \cii\ emission, provided the CGM was pre-enriched by some other process.
However, given that A1689-zD1 is a sub-L*, pre-mature galaxy at $z=7.13$ whose formation history is relatively short, we assume that the metal-enrichment process is related to on-going activity and focus our discussion on the three scenarios of A) star-formation, E) outflow, and F) tidal disruption.

\subsubsection{Star Formation from Satellite Galaxies}

As discussed in Section~\ref{sec:halo_vs_disk}, we find that the spatially-resolved \Sigcii-\Sigsfr\ relation shows a slight excess of \cii\ emission relative to the local relation at low SFR (at the outskirts of the galaxy).
Such deviation from the local relation suggests that the extended \cii\ emission is unlikely to be explained by individual, low-mass, low-SFR satellite galaxies. 
Moreover, following the mass-metallicity relation \citep[e.g.][]{mannucci_fundamental_2010}, such faint, low-mass satellite galaxies generally have low metallicity. 
At this low-$Z$, the \cii\ line emissivity for a given stellar continuum decreases \citep{vallini_cii_2015, olsen_sigame_2017, katz_nature_2022}, which would in fact have the opposite effect as we observe.

\subsubsection{Outflows}

Another possible scenario for the origin of the \cii\ halo is that high-velocity outflows drive carbon out to $r\sim 10$ kpc. 
These outflows are responsible for the metal enrichment of the CGM, and this interpretation has been supported by observations of ongoing outflows, as traced by the \cii\ line, in high-SFR galaxies \citep{gallerani_alma_2018, ginolfi_alpinealma_2020, herrera-camus_kiloparsec_2021a}.
Since the majority of \cii\ is thought to be radiated from PDRs in the cold, neutral hydrogen gas clouds, the broad wing feature observed in \cii\ suggests that the outflowing gas is dominated by cool, neutral gas. 

However, galactic outflows in general are known to be made up of multiple phases of gas, with the archetypal example of M82 showing a hot phase observed in X-ray emission ($T \gtrsim 10^6$ K), a warm ionized phase ($T \sim 10^4$ K), cold and warm molecular phases and a cold atomic phase \citep{heckman_galactic_2017}.
The multiphase nature of galactic outflows suggests that the \cii\ line may not reliably trace outflows in all galaxies.
In fact, \citet{spilker_ubiquitous_2020} find that molecular outflows are ubiquitous in $z >4$ galaxies, but not always identifiable by \cii\ emission.
We show in Section~\ref{sec:line_profiles} that the \oiii\ line shows significant $\sim 5\sigma$ residuals in the high-velocity region after fitting a single Gaussian model. 
By contrast, we find no significant residuals in \cii. 
This may suggest the presence of ongoing outflows dominated by hot, ionized gas, rather than cool, neutral gas in \zD.

The theoretical connection between these hot outflows and the observed, cool, \cii\ halo depends on the cooling time of the hot ionized gas ($t_{\rm cool}$).
Optically thin intergalactic gas at $z\sim 3.5$ (with temperature $T\sim 10^6$ K and an overdensity $\delta \sim 1$--$10$) is predicted to have a cooling time of $t_{\rm cool} \gtrsim 4$ Gyr, which is longer than the Hubble time at this redshift \citep{madau_early_2001}. 
However, this depends on the specific properties of the gas, e.g.~the density, metallicity, and multiphase structure of the outflows. 
In fact, recent ``cooling outflow'' models implemented into the analytical models of \citet{pizzati_outflows_2020} have successfully reproduced the \cii\ halo out to $r\sim 10$ kpc. 
In their models, hot ionized outflows experience catastrophic cooling shortly after their initial launch. 
Depending on the timescale on which this hot outflowing gas cools, we may or may not expect to observe the broad-wings in \cii\ under this scenario. 
This might contribute to the diversity of the presence or the absence of the \cii\ broad-wing feature in recent ALMA high-$z$ studies \citep[e.g.][]{decarli_alma_2018, gallerani_alma_2018, fujimoto_first_2019, stanley_spectral_2019, ginolfi_alpinealma_2020, bischetti_widespread_2019, novak_no_2020, herrera-camus_kiloparsec_2021a, spilker_ubiquitous_2020, izumi_subaru_2021a, izumi_subaru_2021}.
It is worth noting that in the case of \zD\ the \cii\ line profile may also be composed of narrow+broad components, as with the \oiii\ line, but degenerate with a single Gaussian profile. 
In this scenario, we are observing multi-phase outflows in an early galaxy including both cold neutral and hot ionized gases.

We also note that outflows may also be capable of forming the extended dust structure potentially observed around A1689-zD1 (Section~\ref{sec:morphology}).
Though the outflow may also contribute to destruction of dust, the existence of dust from $\sim 20$ kpc to a few Mpc from galaxies has been confirmed from the reddening measurements of background objects \citep[e.g.,][]{menard_measuring_2010, peek_dust_2015,tumlinson_circumgalactic_2017}.
\citet{fujimoto_alpinealma_2020} report that 1 out of 23 isolated ALPINE galaxies shows a dust continuum profile similar to the \cii\ emission, despite compact UV continuum. 
Interestingly, the object also shows a large Ly$\alpha$ EW even with the large amount of dust, and the authors argue a potential mechanism of outflows dominated by hot ionized gas, which may serve to eject dust out to the CGM, keep the dust hot, and allow for the Ly$\alpha$ line to escape from the system. 
If the hot-mode outflows last only for a short timescale, this extended dust continuum structure could be somewhat rare in the early universe.

\subsubsection{Tidal stripping}

As previously noted, \zD\ is a clumpy object with multiple spatial/spectral components necessary to fit the \cii\ and \oiii\ datacubes \citep[][Knudsen et al.~in prep]{wong_alma_2022}. 
This raises the possibility that \zD\ is actually multiple merging systems, and that the \cii\ halo arises from the dissipation of mechanical energy via turbulence and shocks associated with tidal stripping. 
Such tidal effects would also be expected to eject material from the galaxy, including dust and old stars that would contribute to the ISRF in the CGM.
These effects have been explored by \citet{ginolfi_alpinealma_2020a}, who observe \cii\ emission from circumgalactic gas structure around a massive ($\sim 10^{10} M_\odot$), merging galaxy system at $z=4.57$ in the center of a protocluster environment.

The close projected distances ($\sim 0.5''$) of the different components in \zD\ make difficult a robust distinction between multiple, merging systems and a single complex, clumpy object. 
Moreover, \zD\ is a sub-L*, low-mass ($\sim 10^9 M_\odot$) galaxy, and the galaxy overdensity is not identified around \zD.
As such, the situation may be different from a system of massive merging galaxies in the center of a high redshift protocluster. 
Nevertheless, minor mergers have been predicted to drive significant kinematic disruption of gas in low-mass galaxies \citep[see e.g.~the simulations of ][]{kohandel_velocity_2020}, which could power \cii\ emission.  
While further analysis of the kinematics of the object is needed to explore the merging system scenario (and will be explored in Knudsen et al.~in prep), it should be noted that tidal disruption in a merging system may play a role in producing the \cii\ halo around \zD.

\subsubsection{Summary: origin of the \cii\ halo}

Our observational results suggest the presence of outflows, dominated by hot, ionized gas, in and around the star-forming core of \zD. 
In conjunction with the analytical models of \citet{pizzati_outflows_2020}, our results imply that the \cii\ halo likely formed via some combination of a) hot, ionized outflows and subsequent catastrophic cooling, or b) past outflow activity including the cool, neutral gas.
Our results do not rule out the possibility that the CGM was pre-enriched by past outflow/merger activity and the observed \cii\ halo is powered by strong ionizing photons from the galaxy core or extended \hii\ regions. 
In any case, our results support the connection between the \cii\ halo and the outflow history of the galaxy, and highlight the importance of considering the multiphase nature of galactic outflows.

However, it remains a possibility that the central and secondary components trace a kinematically complex bulge and diffuse disk of the galaxy, respectively, where the latter is missed in previous {\it HST} observations of the rest-frame UV continuum. 
Rest-frame optical observations with \textit{JWST}, which have been scheduled for Cycle 1, will map out the entire stellar distribution and help to answer definitively whether the extended \cii\ emission is part of the galaxy or the CGM.

\subsection{Origin of High \Loiii/\Lcii\ Ratios at $z>6$}\label{sec:ratios}

Recent evidence from ALMA has shown that high redshift galaxies tend to have ($\sim 3$--$10$ times) higher \Loiii/\Lcii\ ratios than local galaxies \citep{harikane_large_2020, hashimoto_dragons_2019, carniani_missing_2020, bakx_alma_2020}. 
This may suggest different properties of the ISM at high-$z$, such as a higher ionization parameter, PDR deficit, higher burstiness parameter \citep{vallini_high_2021}, or lower C$/$O abundance ratio than solar \citep{becker_highredshift_2011, katz_nature_2022}.

Using the best sensitivity \oiii\ and \cii\ maps so far obtained at this redshift, we have shown the existence of the \cii\ halo and \oiii\ outflows in \zD, both of which may be additional factors to explain the high \Loiii/\Lcii\ ratios in $z > 6$ galaxies.  
Although our results rely on one galaxy, A1689-zD1 is just one of the abundant population of sub-L* galaxies whose properties may be a representative picture in the Epoch of Reionization. 
Therefore, we discuss here the contribution from the \cii\ halo and \oiii\ outflows to the \Loiii/\Lcii\ ratio in \zD.
A full analysis of the spatially-resolved \Loiii/\Lcii\ ratio in \zD\ will be presented in Knudsen et al. (in prep)

\begin{figure}
	\centering
	\includegraphics[width=\linewidth]{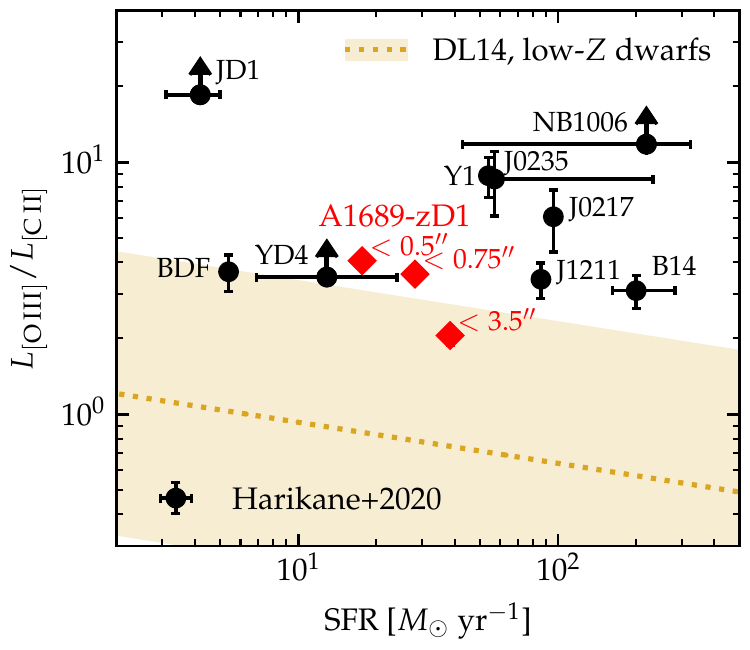}
	\caption{The \Loiii/\Lcii\ ratio vs.~SFR for \zD\ in multiple apertures (red diamonds), compared to the literature sample from \citet[][black circles]{harikane_large_2020}.
	The SFR values are computed as described in Section~\ref{sec:SFR}.
	While the \Loiii/\Lcii\ ratio is lowered by a factor of $\sim 2$ by accounting for the full \cii\ halo, this cannot account for the entirety of the high \Loiii/\Lcii\ ratios in high redshift galaxies.}\label{fig:LOIII_LCII_halo}
\end{figure}

\subsubsection{Contribution from \cii\ halo}

One possible explanation for the high \Loiii/\Lcii\ ratios in $z>6$ galaxies is the systematic existence of, and failure to account for, the extended \cii\ emission on CGM scales.
This has been investigated by \citet{carniani_missing_2020}, who find that correcting for surface brightness dimming in high-resolution, low S/N observations can lower the \Loiii/\Lcii\ ratios by a factor of $\sim 2$.

Figure~\ref{fig:LOIII_LCII_halo} shows the \Loiii/\Lcii\ ratio as a function of SFR for \zD, in multiple circular apertures of different sizes. 
These different apertures each encompass different amounts of the \cii\ halo.  
We additionally show the 9 galaxies studied in \citet{harikane_large_2020}. 
By accounting for the full extent of the \cii\ halo ($< 3.5''$), we find a lower \Loiii/\Lcii\ ratio of $\sim 2$, compared to $\sim 4$ for the central region of the galaxy ($< 0.5''$). 
That is, accounting for extended \cii\ emission can help to reconcile the high \Loiii/\Lcii\ ratios with lower ratios found in local galaxies, but cannot bring these values completely in line with local relations.

\subsubsection{Contribution from \oiii\ outflows}

Another possible explanation for the high \Loiii/\Lcii\ ratio, at least in the case of \zD, is the contribution of the ``broad-wing'' feature in the \oiii\ spectrum to the total \oiii\ luminosity. 
If this emission indeed corresponds to hot, ionized gas in outflows, rather than \hii\ regions around massive stars, we would expect to see a higher \Loiii/\Lcii\ ratio than local universe galaxies not undergoing this outflow. 

\begin{figure}
	\centering
	\includegraphics[width=\linewidth]{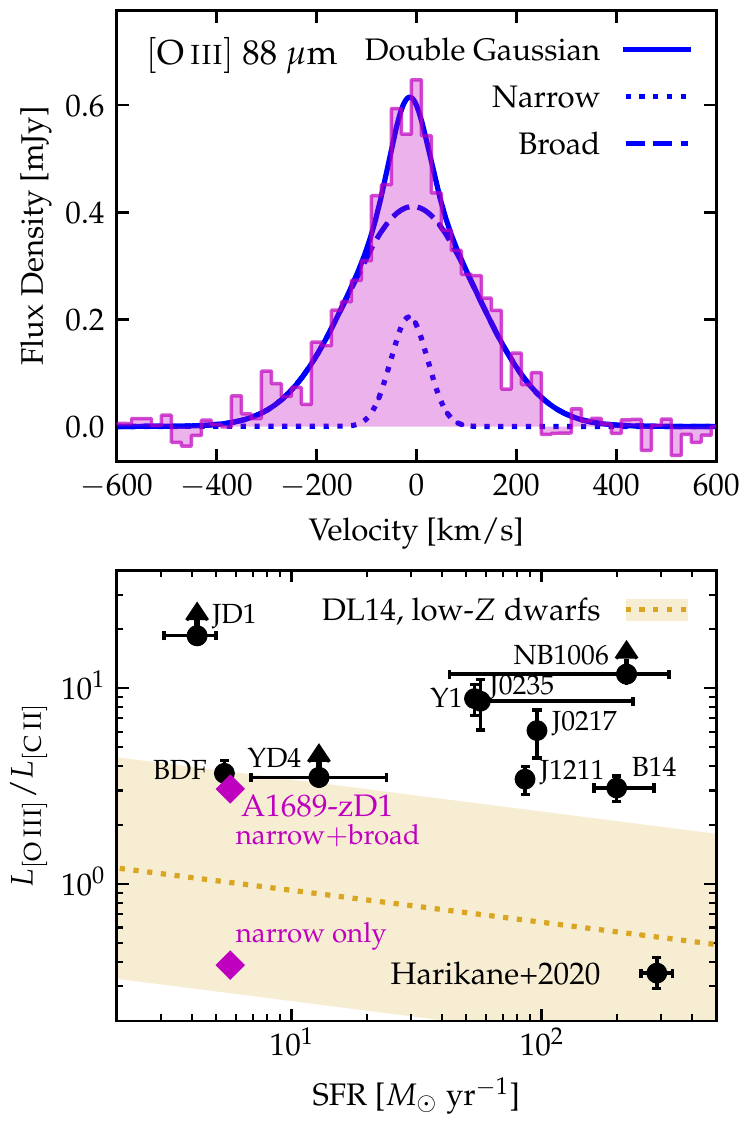}	
	\caption{\textit{Top}: \oiii\ line profile in the central 0.25'' region as in Section~\ref{sec:line_profiles}, but with a double Gaussian fit applied. \textit{Bottom}: The \Loiii/\Lcii\ ratio vs.~SFR for \zD, as in Figure~\ref{fig:LOIII_LCII_halo}, for the total \oiii\ luminosity as well as only the narrow component.}\label{fig:LOIII_LCII_outflow}
\end{figure}

To examine this, we analyze the \oiii\ line profile in \zD. 
Since the single Gaussian fitting in \oiii\ shows significant residuals (Section~\ref{sec:line_profiles}), we perform two component (narrow+broad) Gaussian fitting to the \oiii\ line profile. 
Figure~\ref{fig:LOIII_LCII_outflow} shows the \oiii\ line profile in the same central aperture as Figure~\ref{fig:line_profiles}, with a two-component Gaussian fit applied.
We see that the broad component dominates the overall spectrum, with the narrow component responsible for only about 13\% of the total luminosity.

The bottom panel of Figure~\ref{fig:LOIII_LCII_outflow} shows how the \Loiii/\Lcii\ ratio changes when calculated using the total \oiii\ luminosity (narrow+broad) vs.~the luminosity of only the narrow component. 
We use the same SFR for both points, derived from the UV+FIR emission as described in Section~\ref{sec:SFR}. 
We see that, while this central region shows an overall \Loiii/\Lcii\ ratio of $\sim 3$, the narrow component alone shows a ratio of $\sim 0.4$. 
Assuming the low-$Z$ dwarf galaxies in the \citet{delooze_applicability_2014} sample do not show significant outflows, this result suggests that outflowing gas in the hot, ionized phase could be a factor contributing to the high \Loiii/\Lcii\ ratios in high-$z$ galaxies. 
While the outflow signature in the \oiii\ line is not universally identified around the high-$z$ galaxies with high \Loiii/\Lcii\ ratios, it has been tentatively detected in the galaxy NB1006 \citep{inoue_detection_2016}.
Importantly, the S/N of the \oiii\ line in our data is at least $4$--$8$ times larger than previous studies \citep[e.g.][]{harikane_large_2020}, which could explain the non-detections of the \oiii\ outflow signature.

\section{Summary}\label{sec:summary}

In this paper, we present deep ALMA observations of the strongly lensed, $z = 7.13$ galaxy \zD, focusing on the \cii\ $158~\mu$m and \oiii\ $88~\mu$m fine-structure lines as well as the FIR continuum. 
Utilizing archival {\it HST} data to probe the UV continuum, we compare the morphology and spatial extent of each wavelength of emission to examine the multiphase nature of the ISM. 
Our major findings are summarized below. 
\begin{enumerate}
	\item We detect both \cii\ and \oiii\ emission at high S/N, as well as dust continuum emission in both bands. The \cii\ emission is extended out to $r \sim 12$ kpc from the galaxy center, while the \oiii\ and UV continuum emission extend only to $r \sim 4$ kpc.
	\item Despite low S/N, we find that the dust continuum follows a similar profile to the \cii\ line and is more extended than the \oiii\ line or UV continuum.
	\item Utilizing multiwavelength FIR continuum data, we measure a dust temperature of $T_{\rm dust} = 41^{+17}_{-14}$ with a dust emissivity index of $\beta = 1.7^{+1.1}_{-0.7}$. We find slight ($\sim 1$--$2\sigma$) differences in the dust continuum radial profiles, indicating that dust temperature might vary in space; in fact, we find tentative indications that the temperature decreases from $\sim 50$ K at the galaxy center to $\sim 30$ K at $r\sim 1.5''$.
	\item The spatially-resolved \Sigoiii-\Sigsfr\ relation shows an excess of \oiii\ emission in the galaxy core as compared to local dwarf galaxies. By contrast, the \Sigcii-\Sigsfr\ relations shows a slight deficit of \cii\ emission in the galaxy core and a slight excess in the outskirts, consistent with $z \sim 2$--$4$ SFGs.  
	\item The central, high-$\Sigma_{\rm SFR}$ region shows a \cii\ line width of $252 \pm 21$ km/s and an \oiii\ line width of $182 \pm 31$ km/s, indicative of the larger spatial extent of the \cii\ line along the line of sight. We find evidence for ongoing hot ionized outflows traced by \oiii, with $\sim 5\sigma$ residuals from a single Gaussian model at $\pm 100$-$400$ km/s. This suggests a possible origin of the \cii\ halo from hot ionized outflows and subsequent cooling.  
\end{enumerate}
Much remains uncertain about the ISM/CGM properties, kinematics, and evolutionary stage of \zD. 
As \zD\ will be observed in {\it JWST} Cycle 1 (Prop. ID 1840, PI: Alvarez-Marquez) future work will be able to take full advantage of imaging from the rest-frame UV to the optical.
The strongly gravitationally lensed nature of this galaxy makes it a prime target with which to constrain the processes of early galaxy evolution.

\acknowledgments

The authors are grateful to Francesca Rizzo, Andrea Pallottini, Livia Vallini, Mahsa Kohandel, Andrea Ferrara, David Frayer, and Harley Katz for helpful discussions regarding the physical interpretations of the extended \cii\ emission around \zD. 
This paper makes use of the following ALMA data: ADS/JAO.ALMA\#2015.1.01406.S, ADS/JAO.ALMA\#2017.1.00775.S. 
ALMA is a partnership of ESO (representing its member states), NSF (USA) and NINS (Japan), together with NRC (Canada), MOST and ASIAA (Taiwan), and KASI (Republic of Korea), in cooperation with the Republic of Chile. 
The Joint ALMA Observatory is operated by ESO, AUI/NRAO and NAOJ.
The National Radio Astronomy Observatory is a facility of the National Science Foundation operated under cooperative agreement by Associated Universities, Inc.
This work was supported by the U.S. NSF under grant OISE-2005578 (DAWN-IRES). 
H.B.A.~acknowledges the support and collaboration of the mentors and fellow students involved in the 2021 DAWN-IRES program. 
S.F.~and D.W.~acknowledge support from the European Research Council (ERC) Consolidator Grant funding scheme (project ConTExt, grant No. 648179) and Independent Research Fund Denmark grant DFF–7014-00017. 
The Cosmic Dawn Center is funded by the Danish National Research Foundation under grant No. 140.
K.K.~acknowledges support from the Knut and Alice Wallenberg Foundation.
T.H.~was supported by Leading Initiative for Excellent Young Researchers, MEXT, Japan (HJH02007) and KAKENHI (20K22358).
A.K.I.~is supported by NAOJ ALMA Scientific Research Grant Code 2020-16B.
M.J.M.~acknowledges the support of the National Science Centre, Poland through the SONATA BIS grant 2018/30/E/ST9/00208.
Much of this work was conducted on the ancestral land of the Meskwaki, Sauk, and Ioway Peoples.

\newpage
\appendix 
\section{Supplementary Figures}\label{appendix:A}

\begin{figure}[h]
    \figurenum{A.1}
    \centering
    \includegraphics[width=0.66\linewidth]{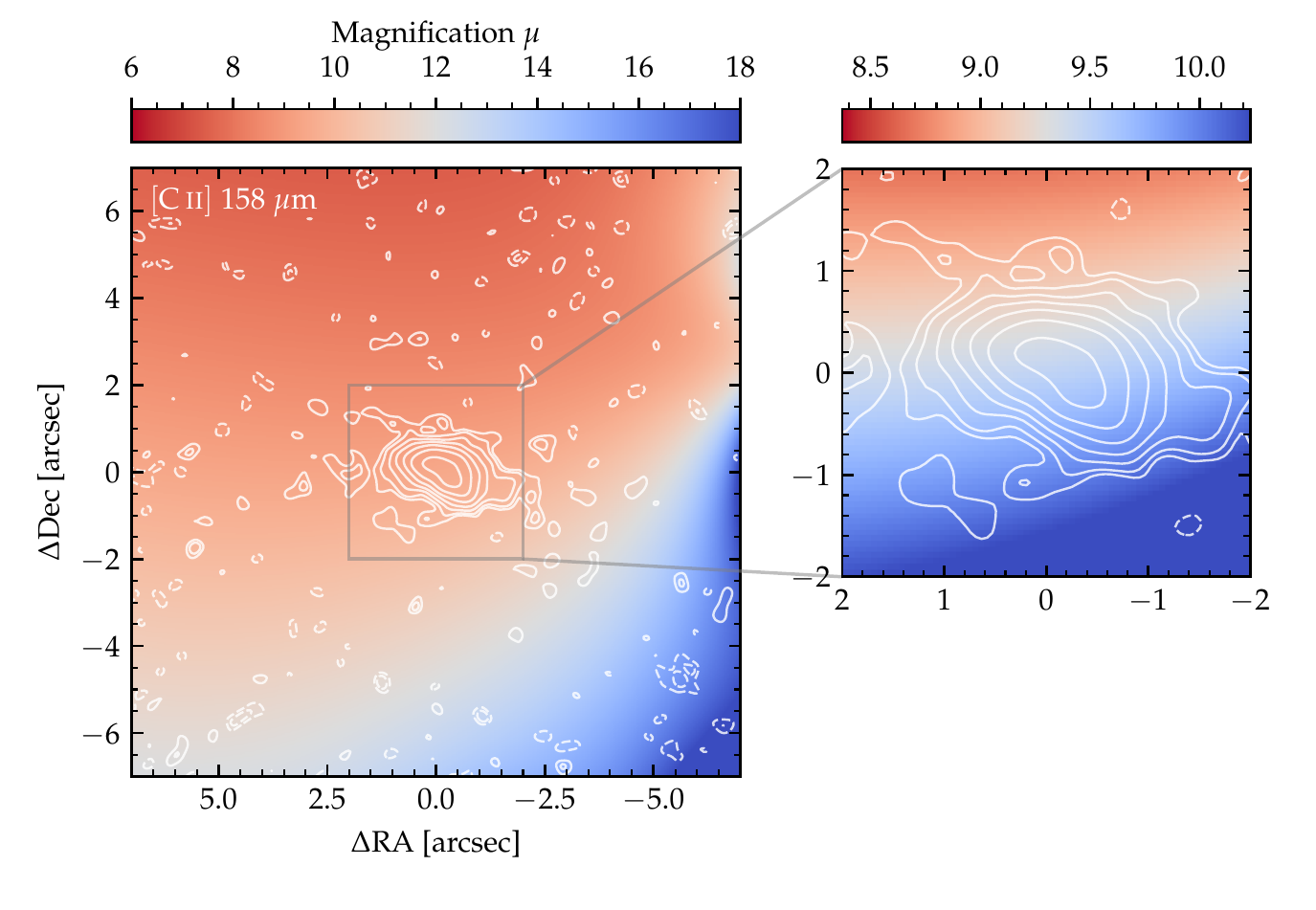}
    \caption{Map of the lensing magnification of A1689-zD1. Contours indicate the \cii~$158~\mu$m line intensity, and the background is colored according to the magnification at that point. 
    The magnification is highest in the southwest corner of the field, but largely uniform (within $\sim 5\%$) over the galaxy.}\label{fig:mag_map}
\end{figure}

Figure~\ref{fig:mag_map} shows a map of the magnification $\mu$ across the field. 
While the magnification is highest in the southwest corner of the field, and lowest in the northeast, it is quite uniform across the source. 

\begin{figure*}[h]
	\figurenum{A.2}
	\centering
	\includegraphics[width=\linewidth]{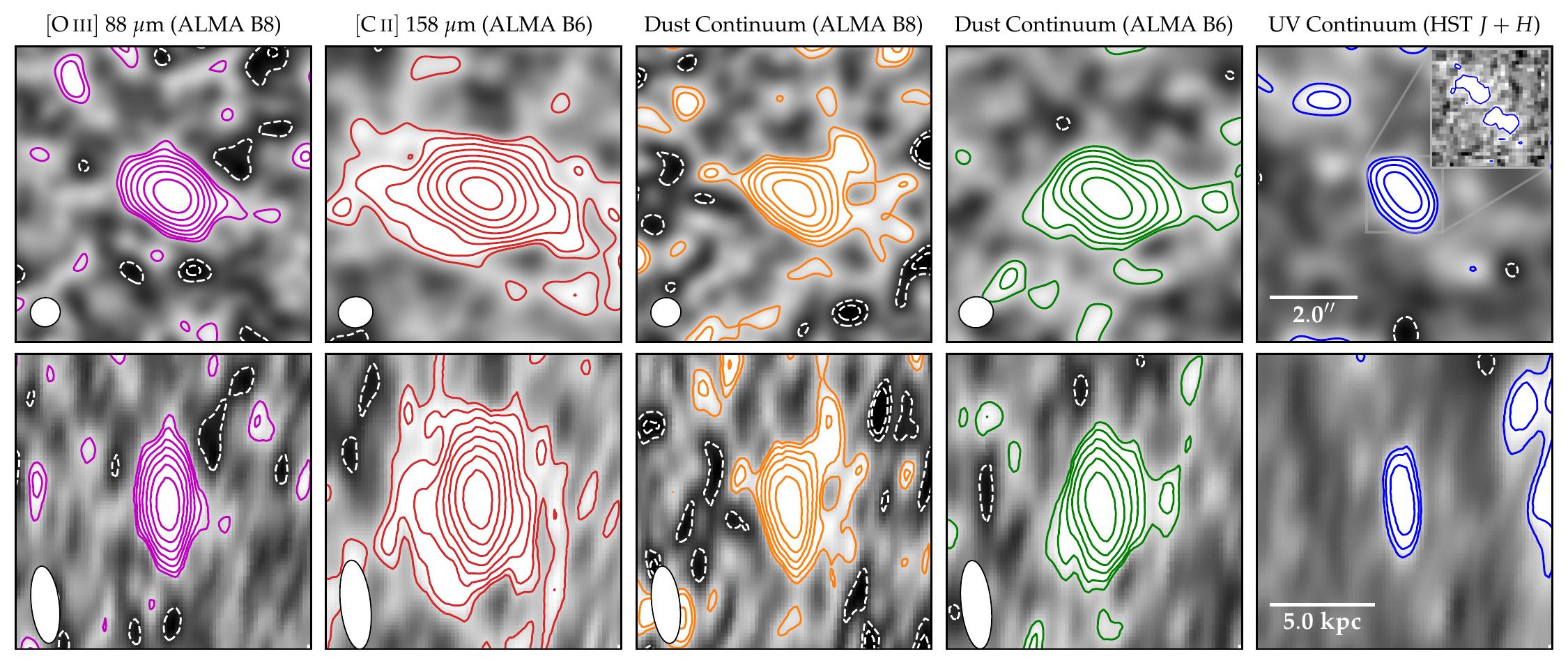}	
	\caption{{\it Top}: Maps of each emission, as in Figure~\ref{fig:maps}, but with a {\it uv}-taper of $0.5''$ ($0.7''$) applied to band 8 (band 6) maps. {\it Bottom}: Maps of each emission reconstructed into the source plane.}\label{fig:maps_uv}
\end{figure*}

We show in Figure~\ref{fig:maps_uv} the 2D maps of each emission, in the image plane (as in Figure~\ref{fig:maps}) as well as reconstructed into the source plane, with a {\it uv}-taper of $0.5''$ ($0.7''$ for band 6) applied.
Much of the analysis in this paper is conducted on these {\it uv}-tapered maps to optimize the S/N of the extended, diffuse emission. 
Moreover, we derive radial profiles, gini coefficients, and the dust temperature profile from the {\it uv}-tapered source plane maps, as they provide a direct measurement of the radial distance, corrected for the lens deflection.

\newpage
\bibliographystyle{aasjournal} 
\bibliography{bibliography}

\end{document}